\documentclass[]{aastex631}

\usepackage{amsmath, amssymb}
\usepackage{graphicx}

\newcommand{\bfk}{{\boldsymbol{k}}}
\newcommand{\bfx}{\boldsymbol{x}} 

\newcommand{\expectation}[1]{\left\langle #1 \right\rangle}
\newcommand{\hatn}{{\boldsymbol{\widehat{n}}}}
\newcommand{\recentchange}[1]{{#1}}

\def\apj{ApJ}
\def\apjl{ApJL}
\def\apjs{ApJS}

\def\nat{Nature}

\def\pasp{PASP}

\def\aap{{A\&A}}
\def\prd{PRD}

\def\prl{PRL}
\def\ssr{Space Science Reviews}
\def\aapr{The Astronomy and Astrophysics Review}

\shorttitle{Outer Solar System spacecraft \& $\mu$Hz gravitational waves}
\shortauthors{McQuinn \& McGrath}

\begin{document}

\title{Outer Solar System spacecraft to probe the $\mu$Hz gravitational wave frontier}

\author[0000-0001-7961-9735]{Matthew McQuinn}
\email{mcquinn@uw.edu}
\affiliation{Department of Astronomy, University of Washington, 3910 15th Ave NE, Seattle, WA 98195, USA}

\author[0000-0002-6155-3501]{Casey McGrath}
\affiliation{Center for Space Sciences and Technology, University of Maryland, Baltimore County, Baltimore, MD 21250, USA}
\affiliation{Center for Research and Exploration in Space Science and Technology II, NASA/GSFC, Greenbelt, MD 20771, USA}
\affiliation{Gravitational Astrophysics Lab, NASA Goddard Space Flight Center, Greenbelt, MD 20771, USA}


\begin{abstract}
The microhertz frequency band of gravitational waves probes the merger of supermassive black holes as well as many other gravitational wave phenomena. However, space-interferometry methods that use test masses would require further development of test-mass isolation systems to detect anticipated astrophysical events. We propose an approach that avoids onboard inertial test masses by situating spacecraft in the low-acceleration environment of the outer Solar System. We show that for Earth-spacecraft and inter-spacecraft distances of $\gtrsim 10\,$AU, the accelerations on the spacecraft would be sufficiently small to potentially achieve gravitational wave sensitivities determined by stochastic gravitational wave backgrounds.  We further argue, for arm lengths of $10-30~$AU and $\sim 10\,$Watt transmissions, that stable phase locks could be achieved with 20\,cm mirrors or $5$\,m radio dishes. 
We discuss designs that send both laser beams and radio waves between the spacecraft, finding that, despite the $\sim10^4\times$ longer wavelengths, even a design with radio transmissions could reach stochastic background-limited sensitivities at $\lesssim 0.3\times 10^{-4}$Hz.  Operating in the radio significantly reduces many spacecraft design tolerances.  Our baseline concepts require two arms to do interferometry.  However, if one spacecraft carries a clock with Allan deviations at $10^4$ seconds of $10^{-17}$, a comparable sensitivity could be achieved with a single arm.  Finally, we discuss the feasibility of achieving similar gravitational wave sensitivities in a `Doppler tracking' configuration where the single arm is anchored to Earth. 
\end{abstract}

\section{Introduction}
\label{sec:intro}

It has been less than a decade since the first direct detection of gravitational waves by the LIGO/Virgo collaboration \citep{PhysRevLett.116.061102}.  In subsequent years, the LIGO/Virgo collaboration has cataloged more than a hundred black hole merger events at kilohertz frequencies \citep{2023PhRvX..13d1039A}, as well as several neutron star merger candidates, including the famous 2017 multi-messenger event \citep{2017PhRvL.119p1101A}.  Recently, the observed spectral range of gravitational waves has been extended to almost a nanohertz with the likely detection of a stochastic gravitational wave background using pulsar timing arrays \citep{2023ApJ...951L...6R, Agazie_2023, 2023A&A...678A..50E}, a signal that probably owes to the inspirals of the most massive supermassive black hole binaries.  

A gravitational wave interferometer sensitive to significantly lower frequencies than LIGO/Virgo requires going to outer space because of seismic noise as well as other terrestrial noise sources. The Laser Interferometer Space Antenna (LISA), scheduled for launch starting in 2035 \citep{LISA}, aims to fill in the $10^{-4}-1~$Hz waveband that is intermediate between the pulsar timing arrays and the ground-based efforts like LIGO/Virgo.  LISA will send laser beams between three spacecraft in a triangle configuration with side lengths of $0.017~$AU.  The lasers will work as multiple Michelson-like interferometers, with the aim of measuring phase changes that result from displacements as small as an angstrom.  The reference for these precise displacement measurements must be sufficiently isolated from sources of acceleration (such as the Sun's irradiance variations) to reach the sensitivities needed to detect known astrophysical gravitational wave sources.  Each LISA spacecraft employs the most sensitive accelerometer ever built, which works by monitoring a nearly drag-free test mass. 

Considerable effort has been directed toward finding detection strategies in other regions of the gravitational wave spectrum.  The most exciting frontiers are the decihertz region, between the waveband probed by LISA and LIGO/Virgo \citep{2020CQGra..37u5011A}, 
and the ``microhertz'' band of $10^{-7}-10^{-4}$Hz, which falls between the pulsar timing arrays and LISA.  This $10^{-7}-10^{-4}$Hz band probes the early inspiral of the $\sim 10^{5} - 10^{6.5} M_\odot$ black holes that LISA observes nearer to merger, as well as the inspiral and merger of $10^{6.5}-10^{10} M_\odot$ black holes -- the class of black holes that are associated with quasars and may be more likely to yield an electromagnetic counterpart.  There are a host of other astrophysical sources that fall in the $10^{-7}-10^{-4}$Hz band \citep[e.g.][]{2021ExA....51.1333S}. 
While many proposed methods in this waveband lack sufficient sensitivity for known astrophysical processes, they may still detect larger backgrounds, such as those produced in the early universe \citep{PhysRevD.103.L041302, 2023JHEP...12..194B}.  One idea is to use measurements of the lunar orbit by future laser ranging \citep{2022PhRvL.128j1103B}.  Another is to use the very precise angular localizations of stars to constrain angular variations from passing gravitational waves, i.e. gravitational wave astrometry \citep{2022PhRvD.106h4006W, 2022PhRvD.106b3002F, astrometryGW}.  The proposals forecast to be the most sensitive follow in the spirit of an expanded LISA, where the three spacecraft are situated in an equilateral triangle tracing Earth's orbit  \citep{Folkner2011, 2010MPLA...25..922N}. Two recent examples are the $\mu$Ares and LISAmax concepts \citep{2021ExA....51.1333S, 2023CQGra..40s5022M}. 

The $\mu$Ares concept assumes acceleration isolation to $10^{-15}$~m~s$^{-2}$~Hz$^{-1/2}$ over its proposed frequency band of $10^{-7}-1~$Hz, which contrasts with the LISA acceleration allowance of $10^{-14}$m s$^{-2}$~Hz$^{-1/2}$ at the bottom of the LISA band of $10^{-4}$Hz \citep{LISA}.  LISAmax more conservatively takes the same acceleration control specifications as the LISA mission, allowing it to achieve $100\times$ improved sensitivity over LISA owing to the longer arms. LISAmax additionally extrapolates the LISA acceleration control below the LISA band to $1~\mu$Hz assuming the square root of its error power spectrum scales as $f^{-2}$ \citep{2023CQGra..40s5022M}.  However, there are some acceleration sources for the LISA accelerometer that become important at $\sim 2\times 10^{-5}$Hz and that scale much more strongly than $f^{-2}$ to lower frequencies, such as thermal-mechanical noise \citep[e.g.][]{2019BAAS...51g.243M}. 
Concerned that substantial development in acceleration control would be required for space-interferometers to probe the $\mu$Hz band, \citet{2022PhRvD.105j3018F} considered the possibility of instead establishing stations on two asteroids with orbits around $1$~AU and carefully measuring their relative distance.  Because of their large masses, the asteroids would behave as excellent test masses, avoiding the need for precise acceleration control.

Here we consider another method to avoid onboard acceleration monitoring -- employing spacecraft farther out in the Solar System, reaching distances and inter-spacecraft separations of tens of astronomical units. Abandoning drag-free control was considered by \citet{McKenzie2011} and \citet{Folkner2011} in the context of a LISA-like mission.  The outer Solar System application we consider results in a potentially massive reduction in acceleration sources, as the solar irradiance variations and the solar wind density fall off as $r^{-2}$ with distance from the Sun.  Arms over which the gravitational wave signal is measured can be oriented perpendicular to the spacecraft-Sun direction to further suppress these largely radial accelerations \citep{McKenzie2011}.  
Finally, the longer baselines of our outer Solar System concepts relax the requirements on other system specifications to achieve the same sensitivity to gravitational waves. 

A drawback of such long baselines is that the electromagnetic transmissions between spacecraft would be weak.  However, we argue, that even for spacecraft that are separated by several tens of astronomical units, Watt-scale electromagnetic transmissions are still sufficiently strong to achieve stable phase locks.  Another concern is that only meager $\sim 10\,$kbps downlinks have been achieved to spacecraft in the outer Solar System. Fortunately, only a single phase measurement for every hour of data may be required because of the low frequencies of interest, such that an hour per month of $\sim 10\,$kbps downlinks would likely be sufficient. 

This paper also considers an additional optimization, using radio dishes rather than lasers to measure spacecraft separations.  One difficulty with using lasers pertains to the spacecraft relative velocities: larger relative velocities mean larger differences in the interfering frequencies, $\Delta f$.  As phase errors scale with timing errors $\delta t$ as $\delta \phi \sim \Delta f \delta t$, many of the spacecraft design tolerances would be set by the magnitude of this frequency difference.  While the changes in velocity may be slow enough that small frequency differentials can be achieved by periodically tuning the frequencies with small adjustments to the laser cavity properties, the radio avoids this difficulty by directly measuring the phase of the inter-arm transmissions.  Other advantages of the radio include being insensitive to intensity variations in the transmission as well as relaxed pointing requirements.  However, using radio broadcasts rather than lasers is potentially much less sensitive to gravitational waves due to the $\sim 10^4$ times longer wavelengths.  We show that a radio instrument can still be sufficiently sensitive that acceleration noise (which is insensitive to the transmission wavelength) is dominant over much of the gravitational waveband of interest.  
Another concern with a radio effort is that plasma dispersion would contaminate the measured phases.  We show that this noise can be essentially eliminated using two frequency channels. 

A radio design may be more easily added to other outer Solar System spacecraft, which often already include a relatively large high-gain antenna for telemetry.\footnote{The radio-dish designs we study here could plausibly also execute $\sim 10\;$AU very long baseline interferometry to radio sources (in particular, fast radio bursts), as has been proposed for measuring cosmic distances and dark matter structure in \citet{2023ApJ...947L..23B} and \citet{2024PhRvD.110b3516X}.  Additionally, they may also be able to use the timing of fast radio bursts repetitions to constrain the $\mu$Hertz stochastic gravitational wave background \citep{2024arXiv240712920L}.}  Indeed, there is a long history of using radio broadcasts to track spacecraft velocities and probe gravitational waves \citep[for a review see][]{dopplertracking}.  
There has been recent interest in Doppler tracking in the context of a future outer Solar System mission, where it has been suggested that a large improvement in sensitivity may be possible, pushing Doppler tracking into a regime where it can detect anticipated astrophysical sources \citep{2024arXiv240602306Z}.  To reach the sensitivity benchmarks in \citet{2024arXiv240602306Z} at $f\lesssim 10^{-6}$Hz, we show that a Doppler tracking mission would likely require onboard instrumentation that corrects for the accelerations from solar radiation and the interplanetary plasma. 

\begin{figure}[htbp]
\centering
\includegraphics[width=.9\textwidth]{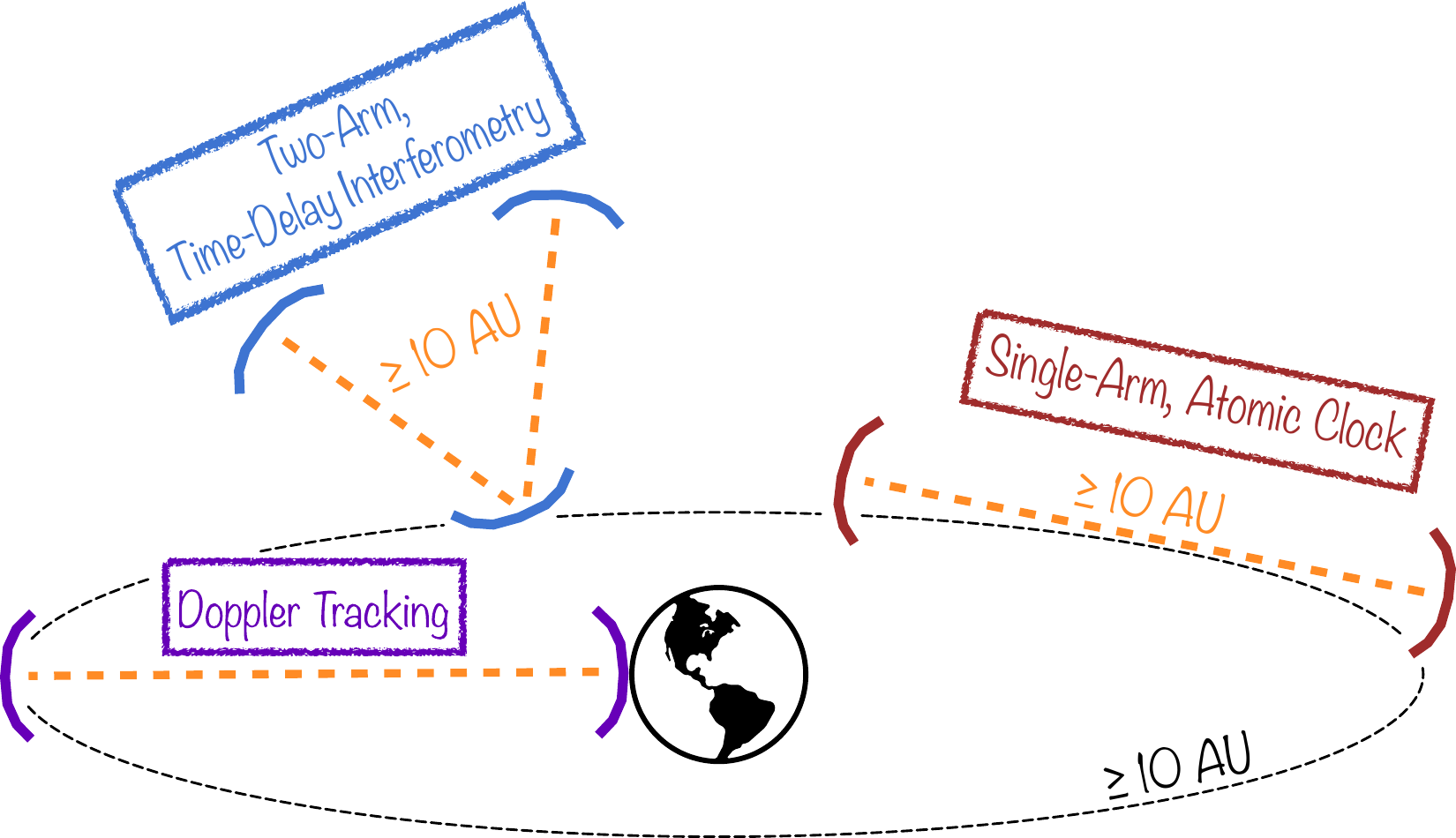}
\caption{
Illustration of the three different architectures considered in this paper: (1) A two-arm design that relies on time-delay interferometry in blue, (2) a single-arm design that relies on a precise atomic clock in red, and (3) an Earth-anchored Doppler tracking design in purple.  Our primary focus is on the two-arm time-delay interferometry concepts, but we discuss the clock requirements to achieve a comparable sensitivity with a single arm.  We contrast the sensitivity of these two designs with the traditional Doppler tracking to outer Solar System spacecraft in \S~\ref{sec:doppler}.   For the calculations in this paper, we consider designs with spacecraft-spacecraft and spacecraft-Earth separations of 10 AU and 30 AU, although the spacecraft would be most easily placed on trajectories where they drift outward rather than orbit the Sun (so that their separations would gradually increase with time).   As a conceptual reference, Saturn orbits between approximately 9-10 AU, Uranus between approximately 18-20 AU, and Pluto between approximately 30-50 AU.}
\label{fig:configurations}
\end{figure}

This paper is organized as follows. Section~\ref{sec:noisesources} discusses the radiometer and acceleration noise sources that are likely to shape the sensitivity of the proposed concepts. Section~\ref{sec:strainpower} uses these estimates to predict the concepts' gravitational wave sensitivity, where we consider the three general mission architectures illustrated in Figure~\ref{fig:configurations}.  Section~\ref{sec:control} elaborates on some of the instrumental considerations that are most relevant.  The appendices discuss the effects of the interplanetary plasma on the phase timing of radio waves, considering dispersion (\ref{sec:dispersion}) and refraction (\ref{sec:refraction}), and they also consider the downlink data rates that these concepts would require (\ref{sec:datarates}), as well as an estimate for their angular resolution (\ref{ap:angres}). 

Unless stated otherwise, 1D power spectra are always half-bandwidth power spectra.  As both electromagnetic and gravitational signal frequencies appear in our calculations, to distinguish them, we generally use wavelengths when referring to electromagnetic transmissions that are sent along arms, and we generally use frequencies when referring to gravitational wave signals and their potential noise sources, being more explicit in our notation in the few cases where we do not. We use Gaussian conventions for electromagnetic quantities such as electron charge (although we do use volt rather than statvolt when referring to spacecraft voltages). 

\section{Strain noise sources}
\label{sec:noisesources}

This section considers the different noise sources that set the gravitational wave sensitivity.  We first consider clock errors and their avoidance through time-delay interferometry (\S~\ref{ss:clock}), then discuss errors related to the strength of the electromagnetic beams (\S~\ref{ss:shot}), and finally discuss acceleration errors (\S~\ref{ss:acc}).  These noise sources are then used to calculate the gravitational wave sensitivity of our concepts in \S~\ref{sec:strainpower}.

\subsection{Clock noise\footnote{\recentchange{`Clock noise' in this section is used in a different manner than in the space laser interferometry literature (such as pertaining to the LISA mission), where this term means the timing jitter from the interplay between the clock noise and spacecraft Doppler shifts \S~\ref{ss:laserdesign}.}\label{footnote:phase}} and its mitigation}
\label{ss:clock}

\paragraph{\textbf{Single arm and an atomic clock}}  Let us first consider a one-arm configuration in which a monochromatic light wave with phase $\phi_{\rm os}$ tied to an onboard oscillator or atomic clock. This wave is sent from one spacecraft to another a distance $L_1$ away and then returns.  If the phase of the incoming signal is compared with the phase on board, up to an overall constant, the phase difference is given by \citep[e.g.][]{1975GReGr...6..439E}
\begin{equation}
\phi_1 = \phi_{\rm os}(t - 2 L_1/c)  - \phi_{\rm os}(t) + \frac{2\pi}{\lambda} L_1 h_1(t) +  \phi_{\rm 1, N},
\label{eqn:phi1}
\end{equation}
 where ${2\pi}c/{\lambda}$ is the electromagnetic wave's angular frequency, $\phi_{\rm 1, N}$ is the noise, and ${2\pi}/{\lambda} \times L_1 h_1(t)$ is the gravitational wave contribution to the phase in the long wavelength limit $c/f \gg L_1$. (This limit applies at our target of $f=1~\mu$Hz, where the gravitational wavelength is $2000\text{ AU}$.)  The phase of the electromagnetic signal $\phi_{\rm os}$ has noise that can be related to the Allan deviation $\sigma_y(\tau)$ of the clock. 
 If we assume that the clock's frequency noise is white, such that $\sigma_y(\tau)^2 \propto \tau^{-1}$, the phase noise power is \citep{IEEEclock} 
\begin{equation}
S_{{\rm os}}(f) \equiv 2 \, T^{-1}\expectation{\widetilde{\phi}_{\rm os}^2} =  \frac{2  \, ({2\pi}c/{\lambda})^2 \times \tau \times \sigma_y(\tau)^2}{(2 \pi f)^2},
\label{eqn:Sphiclock}
\end{equation}
where $\widetilde{\phi}_{\rm os}$ is the Fourier transform of $\phi_{\rm os}(t)$ over the time interval $T$.
Since the phase difference given by equation~(\ref{eqn:phi1}) involves $\phi_{\rm os}$ at two times, the phase noise power relevant for constraining $h_1$ is 
\begin{equation}
S_{\rm os, tot} = 2S_{{\rm os}}\left[1- \cos(4 \pi L_1 f/c) \right]. 
\label{eqn:phiclock}
\end{equation}
Equation~(\ref{eqn:phiclock}) allows us to calculate the long-wavelength strain noise power of our interferometer owing to white frequency modulation clock noise:
\begin{equation}
\widetilde{h}_{\rm os}^{\rm LW} = ~~\frac{\sqrt{ S_{{\rm os}, \rm tot}}}{({2\pi}/{\lambda}) L_1} = {2^{3/2} ~  (1~ \rm s)^{1/2}} \times   \sigma_y(1~ \rm s) =  2.8\times10^{-13} ~{\rm Hz}^{-1/2} \left( \frac{\sigma_y(1~ \rm s)}{10^{-13}}  \right),
\label{eqn:atomic_clock}
\end{equation}
where we use the convention of evaluating the Allan deviation at a second. 

The Deep Space Atomic Clock -- an atomic clock launched in 2019 and a prototype for future interplanetary space missions -- achieved $\sigma_y(\tau) \approx 2\times 10^{-12} (\tau /1 ~{\rm s})^{-1/2}$ for $\tau \lesssim 10^5~$s in a 2019 launch to a geostationary orbit \citep{Burt2020}.  The outer Solar System would avoid Earth magnetic field variations and the $9^\circ$ temperature variations that it experienced, perhaps allowing the Deep Space Atomic Clock to achieve a noise level closer to its $10\times$ improved performance in the laboratory \citep{Burt2020}, a precision that has also been achieved by other space-certified clocks \citep{Wang_2021}. The best atomic clocks on Earth have achieved $\sigma_y$ values of $\mathcal{O}\left(10^{-19}\right)$ \citep{greatatomicclocks, 2024PhRvL.133b3401A}, and some studies have considered space-based gravitational wave detectors with atomic clocks of comparable precision \citep{2015arXiv150100996L, PhysRevD.94.124043}.  

The strain sensitivity given by equation~(\ref{eqn:atomic_clock}) for timing precision of the best space qualified clocks is close to the sensitivity to detect the strain from equal mass $\gtrsim 10^8M_\odot$ supermassive black hole mergers (as will be discussed in \S~\ref{sec:strainpower}).   However, a couple of orders of magnitude more precise atomic clocks are likely required to reach the strain sensitivities that could justify a spacecraft relying on atomic clocks. Due to the rapid development of terrestrial atomic clocks, a one-arm mission that relies on an atomic clock may become a possibility in the near future, and such a mission would be most motivated at $f\sim 1\mu$Hz owing to the larger strain amplitudes from anticipated astrophysical sources.  We provide estimates for the mission sensitivity for different $\sigma_y$ in \S~\ref{sec:strainpower}.  We next review time-delay interferometry, an approach that essentially eliminates clock noise.

\paragraph{\textbf{Two-arm and time-delay interferometry}} The clock noise can be greatly reduced with a $\geq 2$-arm interferometer using a technique called time-delay interferometry \citep{1985ESASP.226..157F, TDIreview}.  We consider a two-arm configuration with a home spacecraft broadcasting monochromatic waves (either with lasers or high-gain antennas) to two other spacecraft, with these spacecraft a distance $L_1$ and $L_2$ away.  This wave travels to the other spacecraft and then is sent back and compared to the reference wave at the home spacecraft. The phase differences measured from each of the two arms are
\begin{equation}
\phi_1 = \phi_{\rm os}(t - 2 L_1/c)  - \phi_{\rm os}(t) + \frac{2\pi}{\lambda} L_1 h_1(t) +  \phi_{\rm 1, N}; ~~~\phi_2 = \phi_{\rm os}(t - 2 L_2/c) - \phi_{\rm os}(t) + \frac{2\pi}{\lambda} L_2 h_2(t) + \phi_{\rm 2, N},
\end{equation}
following the same conventions as in equation~(\ref{eqn:phi1}). The data from a three-spacecraft configuration linked over two arms can be synthetically combined to create the following `time-delay' observable:
\begin{equation}
X(t) = [\phi_2(t- 2L_1/c) - \phi_2(t)] -  [\phi_1(t- 2L_2/c) - \phi_1(t)]. 
\label{eqn:Xt}
\end{equation}
Remarkably, the clock noise -- $\phi_{\rm os}$ -- cancels in this expression -- and, when considering a single frequency $f$ and the limit $L_1= L_2$, other noise sources are suppressed by the same factor as $h_1$ in this estimator -- meaning that $X(t)$'s sensitivity to $h_1$ is the same as that of $\phi_1 - \phi_2$ but without clock noise.\footnote{In practice, the arm lengths are not perfectly known, leading to clock noise not perfectly canceling in the time delay observable \citep{Barke2015}.  Fortunately, this leads to the RMS strain error from clock noise being suppressed by the factor $\sim \delta L/L$ compared to in equation~(\ref{eqn:atomic_clock}), where $\delta L$ is the uncertainty in the arm lengths.  If the positions of the detectors are known to tens of centimeter precision, as is reasonable for ranging on Solar System scales \citep{2023ApJ...947L..23B}, this leads to a negligible phase error in our calculations. Additionally, equation~(\ref{eqn:Xt}) is one example of a synthetic time-delay interferometry observable that can be constructed. More advanced combinations can be created that can, for example, further suppress the clock noise even when the arm lengths are changing during the measurement~\citep[see, for example,][]{TDI_with_motion}. 
}  For our calculations, the knowledge that $X(t)$ exists allows us to consider the interferometric observable $\phi_2 - \phi_1$ in the absence of clock noise.\footnote{For the case $L_1 < L_2$ and the long-wavelength limit $c/f \gg L_1$, the noise of this time-delay interferometry observable is increased by $L_2/L_1$ relative to what we would compute from the observable $\phi_1 - \phi_2$ without clock noise and in which both arms have length $L_2$ \citep{PhysRevD.66.062001}.}    

\subsection{Radiometer noise}
\label{ss:shot}

When limited by the strength of the incoming transmission, and deferring some complications that arise for the case of heterodyne interferometry of lasers, a spacecraft can measure the incoming phase using a phase-lock loop with a half-bandwidth error power spectrum of \citep[e.g.][]{misraenge}
\begin{equation}
S_n(k) \equiv 2 \, T^{-1} \expectation{|\widetilde{\phi}_{1, N}|^2} = (C/N_0)^{-1},
\label{eqn:phiCN}
\end{equation}
where $C/N_0$ is the ratio of the carrier power to the unit-frequency noise power. Tildes denote the Fourier dual such that $\widetilde{\phi_N}$ is the Fourier dual of the phase noise in equation~(\ref{eqn:phi1}). Equation~(\ref{eqn:phiCN}) holds for both an interferometric setup using optical lasers or one with radio dishes, although what sets the noise is different between the two cases.  

In the optical, $C/N_0$ is set by the shot noise of the received laser \citep{Barke2015}:  
\begin{eqnarray}
[C/N_0]_{\rm dB-Hz}^\text{Optical} &=& 10 \log_{10}  \frac{ P_{\rm rec}}{2 \eta h c/\lambda}, \nonumber \\
          & =& 33 \, \text{dB-Hz} + 10 \log_{10}\left(\frac{P_{\rm em}}{10 {\rm~W}} \right) + 40 \log_{10}\left(\frac{D}{20 {\rm~cm}} \right) - 10 \log_{10} \eta \nonumber \\
    &&- 20\log_{10}\left( \frac{L}{30 {\rm~AU}} \right) - 10 \log_{10} \left(\frac{\lambda}{1 \mu{\rm m}}\right),\label{dbHzlaser}
\end{eqnarray}
where $ P_{\rm rec}$ is the received power of a Gaussian laser optimized to maximize the received power, $\eta$ accounts for the efficiency of the photodiodes and the fraction of the laser power at the heterodyne frequency, $D$ is the mirror diameter, and we are using the engineering convention of characterizing $C/N_0$ in dB-Hz with the notation $[C/N_0]_{\rm dB-Hz} \equiv 10 \log_{10}(C/N_0)$.  This paper considers $L=10$ and $30$AU, and a LISA-like $\lambda=1\;\mu$m. 

In the radio, the noise is characterized by the effective temperature of the system, $T_{\rm sys}$, and is given by
\begin{eqnarray}
[C/N_0]_{\rm dB-Hz}^\text{Radio}  &=&  10 \log_{10}\left(\frac{P_{\rm rec}}{kT_{\rm sys}} \right), \nonumber \\
    &=& 34 \, \text{dB-Hz} + 10 \log_{10}\left(\frac{P_{\rm em}}{10 {\rm~W}} \right) + 40 \log_{10}\left(\frac{D_{\rm eff}}{5 {\rm~m}} \right) - 10 \log_{10} \left( \frac{T_{\rm sys}}{50{\rm ~K}} \right) \nonumber\\
    &&- 20\log_{10}\left( \frac{L}{30 {\rm~AU}} \right) - 20 \log_{10} \left(\frac{\lambda}{1 {\rm ~cm}}\right), \label{dbHzradio}
\end{eqnarray}
where $P_{\rm em}$ is the emitted power and $D_{\rm eff}$ is the effective diameter of the radio dishes \citep{misraenge}.  We have used the expressions in \citet{2023ApJ...947L..23B} to calculate $P_{\rm rec}$ from $P_{\rm em}$ for a radio dish.  We will use Ka band transmissions with $\lambda=1~$cm for our estimates. For reference, the Voyager and Cassini probes had 4 meter diameter radio dishes, the SMAP spacecraft employed a fold-out 6 meter dish \citep{SMAP_Handbook}, and the RadioAstron satellite had a fold-out 10-meter dish \citep{2012RSB...343...22K}.   
Additionally, values of the receiver noise temperature of $\sim 50\;$K are typical for narrow band receivers at room temperature, even space-qualified ones \citep{rf_lna_technology_landscape}.\footnote{The low-noise amplifiers can be further positioned on a cold plate on the spacecraft to achieve lower $T_{\rm sys}$.} 

To conceptualize the dB-Hz in equations~(\ref{dbHzlaser}) and (\ref{dbHzradio}), which evaluate to $[C/N_0]_{\rm dB-Hz} \sim 30-50$ for the parameter values discussed in this paper, LISA's pilot tones, whose phase is measured and used to correct timing jitter in the analog-to-digital conversion, have $[C/N_0]_{\rm dB-Hz} =75$ \citep{Barke2014LISAMS}, a full four orders of magnitude larger $C/N_0$ than the signal for fiducial values in equations~(\ref{dbHzlaser}) and (\ref{dbHzradio}).  The GRACE-FO mission's laser lock is able to operate at $[C/N_0]_{\rm dB-Hz} =61$ with minimal cycle slips \citep{Bachman_2017}.    
Another point of reference is the $X$-band downlink of Cassini at $10~$AU to 34\,m Deep Space Network antenna -- used for the most precise Doppler tracking experiment -- had $[C/N_0]_{\rm dB-Hz} \approx 40-50$ \citep{Wang2005}. Finally, the Voyager spacecraft at $140~$AU communicated with $[C/N_0]_{\rm dB-Hz} =30$ when transmitting to a $D=70\,$m Deep Space Network antenna \citep{spacecommunications}. 

Despite our nominal specifications resulting in lower $C/N_0$ than most previous space missions, there has been substantial success at ranging with such weak electromagnetic signals.  Indeed, $[C/N_0]_{\rm dB-Hz}\approx 35$ is known as the {\it acquisition threshold} for a receiver locking onto the Global Positioning System ranging code -- roughly the threshold where a delay-lock loop can acquire the frequency and delay of a signal within $1~$ms by brute force search over a grid of delays and frequencies motivated by typical terrestrial uncertainties \citep{misraenge}. As our concepts' velocities and positions would be extremely well constrained, the threshold for acquisition of a ranging code would be even lower.  (Of course, too small of a $C/N_0]$ would result in cycle slipping in the phase meter as discussed below.)\footnote{
Just like in the global positioning system, the wave sent along each arm would likely be modulated by a pseudo-random code.  Once a delay lock is established, the pseudo-random code decorrelates the signal from contaminating signals, such that we would anticipate phase measurements that are limited by either shot or thermal noise will be possible despite the small $C/N_0$.}

Translating $C/N_0$ to a phase noise via equation~(\ref{eqn:phiCN}) yields
\begin{equation}
\widetilde{\phi}_{\rm rms} = \sqrt{S_n(k)} = 0.01 \times 10^{-\left([C/N_0]_{\rm dB-Hz}-40 \right)/20} ~~\text{rad Hz$^{-1/2}$},
\label{eqn:phirmsshot}
\end{equation}
which must satisfy the requirement that $(T_{\phi}/2)^{-1/2} \widetilde{\phi}_{\rm rms} \lesssim 0.1$ for there to be a phase lock with negligible cycle slipping for typical parameters considered in this paper, where $T_{\phi}$ is the effective averaging-time for the phase measurement made by the phase-lock loop \citep[$2/T_{\phi}$ is the bandwidth of the phase-lock loop;][]{cycleslips, misraenge}. This condition and equation~(\ref{eqn:phirmsshot}) means that our concepts require $T_{\phi} \gtrsim 10^{-2}$s, with the exact value depending on their $C/N_0$.  However, $T_{\phi}$ cannot be longer than the time over which the phase changes by an order one value because of displacements from accelerations, clock drifts, or -- in the case of laser transmissions -- frequency drifts. The mean acceleration from the Sun we find leads to displacements of a waveperiod over $\sim 10 \, (r/30 {\rm AU})~$s for laser transmissions and over a much longer period for the radio ones.  The maximum $T_{\phi}$ may also be set by the clock noise, although we find that this is unlikely to prevent values of $T_{\phi} \sim 10^{-2}$s and even possibly much larger, or the lasers' frequency stability.  Defining $S_f$ to be the laser frequency noise power, we find that $S_f T_{\phi} \lesssim 0.3$ must be satisfied at $f\gtrsim T_{\phi}^{-1}$ to not significantly impact the signal-to-noise of a phase measurement: In a simplified model for the phase lock where the phase is measured over a tophat window of width $T_\phi$, we find the requirement for minimal cycle slips changes to $(T_{\phi}/2)^{-1/2} \widetilde{\phi}_{\rm rms} \exp[S_f T_\phi] \lesssim 0.1$ when measuring the phase from two interfering lasers, both with the same white frequency noise $S_f$.  For $\widetilde{\phi}_{\rm rms} \sim 0.01$ and $T_{\phi} = 10^{-2}$s, this condition requires a factor of $\gtrsim 5$ improvement over the requirement on $\sqrt{S_F}$ of LISA's lasers, which have the requirement $\sqrt{S_f} < 30~\text{Hz}/\text{Hz}^{1/2}$ over pertinent frequencies. Improved frequency noise could potentially be achieved with greater thermal control of the resonant cavity or choosing materials with smaller thermal expansion coefficients \citep{10.1117/12.2691441}.\footnote{In contrast to LISA where this phase noise must be maintained to millihertz frequencies to not compromise gravitational wave science, our concepts' $\sim 10^4 \times$ larger phase noise relaxes the allowance on $S_f$ for $f\ll T_\phi^{-1}$ by this factor (for a fixed spacecraft-spacecraft ranging error), which could provide more flexibility in material choices for the resonant cavity that stabilizes the lasers' frequencies \citep{10.1117/12.2691441}.}  The GRACE-FO mission lasers achieved frequency stability of $\sqrt{S_f} \approx 0.4 (f/1~ \text{Hz})^{-1}~\text{Hz}/\text{Hz}^{1/2}$  \citep{Bachman_2017}, \recentchange{and the thermal noise limit at room temperature is estimated to be $\sqrt{S_f} \approx 0.1 (f/1~ \text{Hz})^{-1/2}~\text{Hz}/\text{Hz}^{1/2}$  \citep{PhysRevLett.93.250602}.}  \recentchange{Our estimates for when a phase lock can be achieved are consistent with recent experiments (simulations) of lasers stabilized to an ultra-stable reference cavity such that $\sqrt{S_f} =2~ (0.1) ~\text{Hz}/\text{Hz}^{1/2}$ are able to achieve phase locking with minimal cycle slips when operating at $[C/N_0]_{\rm dB-Hz} =38~(30)$~dB-Hz \citep{PhysRevLett.131.193804, rs16193598}.}

Translating equation~(\ref{eqn:phirmsshot}) one step further into a displacement noise yields
\begin{equation}
\widetilde{\Delta x}_{\rm rms} = \frac{\lambda \widetilde{\phi}_{\rm rms}}{2\pi} = (\overbrace{1.6\times10^{-3}}^{\lambda =1 \,\text{cm}},~\overbrace{1.6\times10^{-7}}^{\lambda =1 \,\text{$\mu$m}}) ~~~\text{cm~Hz$^{-1/2}$~}  \times 10^{-\left([C/N_0]_{\rm dB-Hz}-40 \right)/20},
\label{eqn:xrms}
\end{equation}
where the two displacement noise values correspond to $\lambda = 1~$cm -- the Ka band radio regularly used for ranging -- $\lambda = 1~\mu$m -- a LISA-like near infrared laser.  For reference, LISA aims to achieve much more precise measurements than our nominal values with $\widetilde{\phi}_{\rm rms} = 9\times 10^{-6} $rad~Hz$^{-1/2}$  and $\widetilde{\Delta x}_{\rm rms} = 15 \times 10^{-10}$~cm~Hz$^{-1/2}$ \citep{LISA}.  

Finally, these estimates translate into a precision for how well the gravitational wave strain can be measured.  The long wavelength strain noise for a single arm is
\begin{eqnarray}
\widetilde{h}_{\rm rms}^{\rm LW} &=& \frac{\sqrt{2} \widetilde{\Delta x}_{\rm rms} }{L}, \nonumber \\
 &=& (\overbrace{5\times10^{-18}}^{\lambda =1 \,\text{cm}},~\overbrace{5\times10^{-22}}^{\lambda =1 \,\text{$\mu$m}}) \text{~~~Hz$^{-1/2}$~}  {\cal A}  \left( \frac{L}{30 {\rm~AU}} \right)^{-1} \times 10^{-\left([C/N_0]_{\rm dB-Hz}-40 \right)/20}.
 \label{eqn:hnoiseshot}
\end{eqnarray}
We will eventually use a transfer function that converts the single-arm noise in the low $f$ limit to a two-arm time-delay interferometry measurement that is applicable at all $f$  (\S~\ref{sec:strainpower}).
The factor of $\sqrt{2}$ in the first expression appears because there are uncorrelated measurements of the phase at both spacecraft in an arm. Here, ${\cal A}$ encapsulates the increase in the phase error from using phase measurements at two wavelengths to eliminate plasma dispersion.  Since plasma dispersion is negligible for lasers, for the laser setup, a single wavelength would be used and ${\cal A} = 1$.  We show in Appendix~\ref{sec:dispersion} that ${\cal A} \approx 1.5 ~(1.8)$ if wavelengths differing by a factor of 4~(2) are used when referenced to the shorter wavelength, assuming $T_{\rm sys}$ and $D_{\rm eff}$ are the same at both wavelengths.  

Another possibility is to use nanosecond laser pulses as a clock, rather than the carrier phase tracking considered so far. Laser pulses would be a noisier alternative for the Solar System-scale baselines we consider with 
\begin{equation}
\widetilde{h}_{\rm rms}^{\rm LW} \Big |_{\rm laser-pulses} \sim 2\times 10^{-14} {\rm ~Hz}^{-1/2} \left( \frac{\lambda}{1 ~\mu\rm  m} \right)^{3/2} \left( \frac{L}{30\, {\rm AU}} \right)^2   \left( \frac{D_1 D_2}{[ 1 \, {\rm m}]^2} \right)^{-3} \left( \frac{P_{\rm em}}{10 \,{\rm W}} \right)^{-3/2},
\label{eqn:pulses}
\end{equation}
where $D_1$ and $D_2$ are the sizes of the mirrors on the emitting and receiving telescopes, and this assumes the laser pulses have a width equal to the time between pulses \citep[cf. their eqn. 95]{2022PhRvD.105j3018F}.  We return to this possibility in \S~\ref{sec:doppler}.
\subsection{Acceleration noise}
\label{ss:acc}

\begin{figure}[htbp]
\centering
\includegraphics[width=1\textwidth]{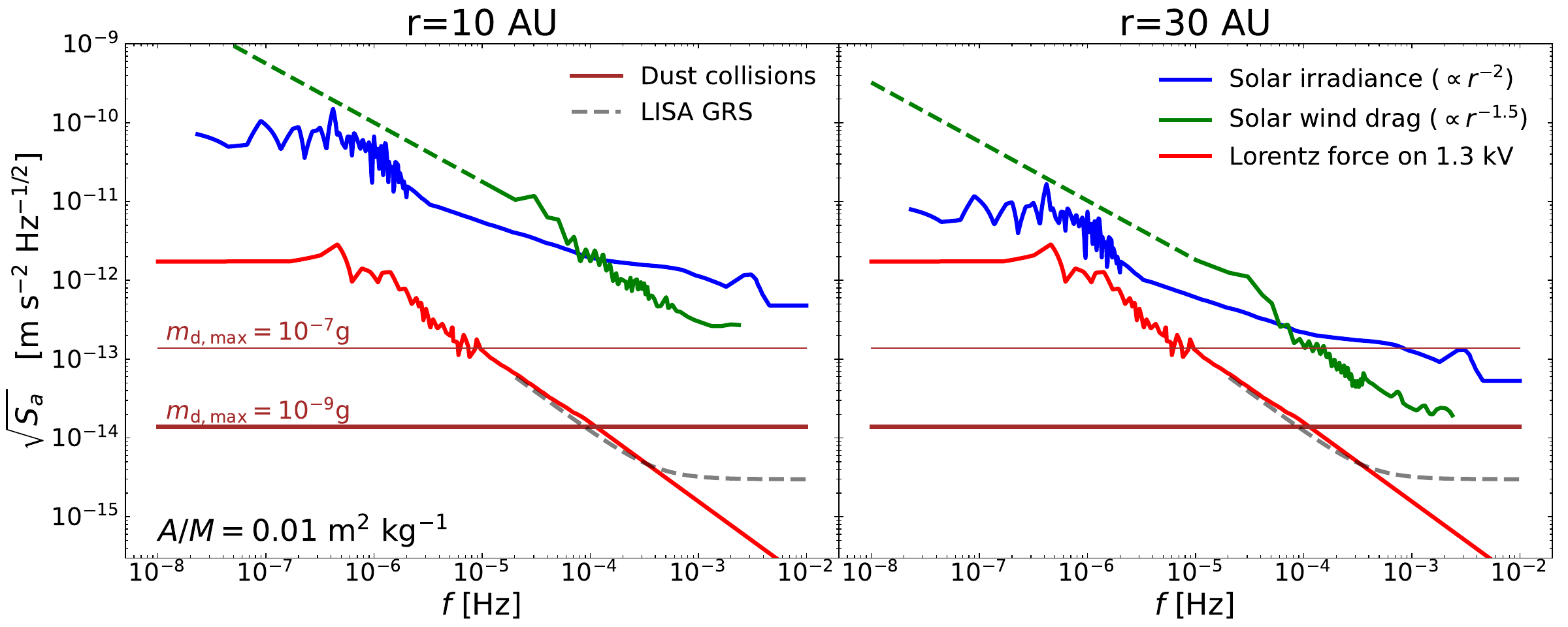}
\caption{Square root of the power spectrum (the amplitude spectral density) of the most important acceleration sources.  
 The calculations assume a spacecraft at a solar distance of $r=10$~AU (left panel) and $r=30~$AU (right panel) with a mass of $M=10^3$kg and an effective area of $A_{\rm eff} =10\;$m$^2$. As irradiance and solar wind are directed radially, a reduction in these forces can be achieved by orienting the arms to be more perpendicular to these radial flows.  Two different maximum dust masses are shown, reflecting uncertainties in the measured distribution.  The contamination of accelerations from $m_{\rm d, max} > 10^{-9}$g grains likely can be cleaned. Also shown is the acceleration goal of the LISA Gravitational Reference Sensor (GRS).}
\label{fig:accelerations}
\end{figure}

In this section, we consider different sources of acceleration on the spacecraft.  We quantify this in terms of the acceleration power spectrum on one detector, defined as $S_a(f) \equiv 2 \, T^{-1}\langle |\widetilde{a}(f)|^2 \rangle$, where $\widetilde{a}(f)$ is the Fourier transform of the acceleration over time $T$. Figure~\ref{fig:accelerations} summarizes our estimates for the important acceleration sources, assuming a fiducial spacecraft effective area over mass of $A_{\rm eff}/M = 0.01\;$m$^2$~kg$^{-1}$ -- further justified  in \S~\ref{ss:twoarmforecasts} -- and spacecraft at a heliocentric radius of $r=10\,$AU (left panel) and $r=30\,$AU (right panel).  The considered acceleration sources are solar irradiance variations, drag from the solar wind, Lorentz forces on the spacecraft assuming the maximum possible spacecraft charge, and dust for two maximum dust masses (as explained later, the lower of the two dust curves is the more applicable).  Figure~\ref{fig:accelerations} also shows the acceleration control specification of the Gravitational Reference Sensor (GRS) on LISA \citep{LISA}. A LISA requirement is to achieve the acceleration control shown by this curve to frequencies as low as $10^{-4}$Hz, with the goal to achieve this to $2\times10^{-5}$Hz.  
This figure shows that at 30~AU the different sources of acceleration are only an order of magnitude larger than the sensitivity of the GRS at $10^{-4}\,$Hz.   Furthermore, since the dominant accelerations are radial with respect to the Sun, additional geometric cancellation is likely when optimizing the spacecraft orientations.  This motivates our overall direction of considering an outer Solar System instrument without a precise accelerometer.

In what follows, we discuss each source of acceleration, ordered roughly by importance.

\subsubsection{Solar irradiance}

During active periods, the Sun shows 0.2\%  peak-to-peak irradiance variations on the timescale of its 27-day rotation period, plus fluctuations over a broad range of timescales.  Figure~\ref{fig:accelerations} shows the power spectrum of irradiance variations measured using the Variability of Solar Irradiance and Gravity Oscillations instrument on the Solar and Heliospheric Observatory \citep[VIRGO/SOHO;][]{2004A&ARv..12..273F}, converting the radiation force to spacecraft acceleration.\footnote{We maintain the major features in the power spectrum using the data from Figure~12 in \citet{2004A&ARv..12..273F}, but we omit much of the fine-grained structure in this power spectrum at $f >10^{-5}$Hz. Note that the effective area used for the conversion from irradiance to acceleration ($A_{\rm eff} =10~$m$^2$) would include the factor-of-two enhancement for a reflective surface directed toward the Sun.} We use measurements over $1996-1997$ that were near the minimum of solar activity \citep{2004A&ARv..12..273F}.   The amplitude of the square root of the irradiance power spectrum varies by a factor of three at $0.1-10\,\mu$Hz over the eleven-year solar cycle, with smaller variations at $>10~\mu$Hz. 


\subsubsection{Solar wind drag}
\label{ss:solarwind}

We find that drag from the solar wind is the other important source of accelerations in addition to irradiance variations.  While the mean drag force scales as $r^{-2}$ with distance from the Sun, the inhomogeneous component falls off somewhat less quickly at $r \gtrsim 10\,$AU owing to the amplitude of the fractional density fluctuations growing with $r$.  From Fourier transforming the electron density time series along the Voyager 2 trajectory, \citet{voyagerturbulence} find a 1D power spectrum of electron density fluctuations with the approximate form  
\begin{equation}
P_e^{\rm 1D} (f, r)\approx P_{e,0}^{\rm 1D} \left( \frac{f}{10^{-5}{\rm Hz}} \right)^{-\beta} \left( \frac{r}{30 \rm AU} \right)^{-3},
\label{eqn:swdensitypower}
\end{equation}
where $ P_{e,0}^{\rm 1D} \approx 10^{11.1}~\text{m$^{-6}$ Hz$^{-1}$}$, and where $-3$ index approximates the radial scaling over $10-50\;$AU.

We can convert from an electron density power to a (radial) spacecraft acceleration power using
\begin{equation}
\sqrt{S_a} =  \mu \, m_p \, v_{\rm sw}^2  A_{\rm eff} M^{-1} \sqrt{P_e^{\rm 1D}(f, r)}.
\label{eqn:Sa}
\end{equation}
 If we take our approximate form for the power given by equation~(\ref{eqn:swdensitypower}), a mean molecular weight of $\mu = 1.2$, and $v_{\rm sw} = 500~$km~s$^{-1}$ for the velocity of the solar wind, we can rewrite equation~(\ref{eqn:Sa}) as
\begin{eqnarray}
\sqrt{S_a} &=&  1.2\times10^{-12} \text{m s$^{-2}$ Hz}^{-1/2}    \left(\frac{A_{\rm eff}/M}{10^{-2} {\rm m^2 \;kg^{-1}}}\right)^{1/2}  \left(\frac{ P_{e,0}^{\rm 1D}}{10^{11.1}~\text{m$^{-6}$ Hz$^{-1}$}} \right)^{1/2}  \nonumber\\
&& \times \left( \frac{f}{10^{-5}{\rm Hz}} \right)^{-\beta/2} \left( \frac{r}{30 \rm AU} \right)^{-1.5}. \label{eqn:dragpl}
\end{eqnarray}

However, rather than assuming some power-law spectrum as in equation~(\ref{eqn:dragpl}), we can compute the drag using the actual electron power that Voyager 2 measured in \citep{voyagerturbulence}.
Figure~\ref{fig:accelerations} shows just this at 30 AU assuming a spacecraft effective area to mass of $A_{\rm eff}/M = 0.01$ m$^{2}$ kg$^{-1}$ and using equation~(\ref{eqn:Sa}).  Figure~\ref{fig:accelerations} illustrates that Voyager 2's electron power spectrum is not exactly a power law. Between $10^{-4}-10^{-2}$ Hz, \citet{voyagerturbulence} find $\beta \approx 2$. At higher frequencies, the spectrum is more consistent with a Kolmogorov turbulence-like $\beta = 5/3$ spectrum.  There is the additional complication that \citet{voyagerturbulence} did not consider $f <10^{-5}$ Hz.  While we think it would be possible to use the Voyager data to probe these lower frequencies, for this study, we extrapolate their measurement with the index $\beta = 1.5$. Extrapolating with a flatter scaling than that observed at higher frequencies is motivated by the power spectrum of the solar wind within several astronomical units, which tends to have a spectral index of $\beta \approx 1$ at these low frequencies. (The properties of the solar wind are likely to be largely maintained as it advects into the outer Solar System.) Additionally, the power spectrum of solar wind density fluctuations is found to be similar to that of magnetic field fluctuations, and the power spectrum of the magnetic field amplitude is observed to flatten at sub-$\mu$Hz frequencies \citep{SMITH1989159}.\footnote{The power spectrum of the magnetic field is proportional to the accelerations from Lorentz forces, which are shown in Figure~\ref{fig:accelerations} and computed using Voyager measurements (cf. \S~\ref{ss:Bfield}).} 

\subsubsection{Dust}

Each collision with a dust grain results in a change of velocity of the spacecraft of $\Delta v_i = m_{d,i} v_{d}/M$, where $m_{d,i}$ is the mass of the $i^{\rm th}$ dust grain and $M$ the mass of the spacecraft. We treat all grains as having the same velocity of $v_{d} = 20~$km~s$^{-1}$, motivated by the interstellar flow of dust that is found to dominate the dust population in the outer Solar System \citep{1994A&A...286..915G}.   If the interferometer arm is not oriented in the flow direction of the interplanetary dust (which has been found to be moving in roughly the ecliptic with longitude $250^\circ$), there is a geometric suppression relative to our estimates.  

The density distribution of interplanetary dust is found to be roughly constant per $\log$ mass between $m_d = 10^{-12}$ and $10^{-9}$g with $d \rho_d/d\ln m_d = 1.2\times10^{-27}$g~cm$^{-3}$ \citep{2000mbos.work...99G}.  The grain size distribution has not been measured above $10^{-9}$g.  We assume $d \rho_d/d\ln m_d = 1.2\times10^{-27}$g~cm$^{-3}$ up to $m_{d,\rm max}$. Since each dust collision can be approximated as causing a step function in the spacecraft velocity, the acceleration of the spacecraft $dV/dt$ is then a sum of Dirac $\delta$-functions at the collision time $t_i$, which in Fourier space is $d\widetilde{V}/dt = \sum_i \Delta v_{d,i} \exp[-i \omega (t - t_i)]$.  If we treat each grain as uncorrelated in time, then the acceleration power spectrum is 
\begin{eqnarray}
\sqrt{S_a} &\equiv & \left(2 T^{-1} \left\langle \left| \frac{d\widetilde V}{dt} \right|^2 \right\rangle \right)^{1/2} = \left( 2\int_0^{m_{\rm d, max}} d\ln m_d \, \times{d\rho_{d}}/{d\ln m_d}  \times \left[\Delta v( m_{d}) \right]^2\right)^{1/2},  \\
 &=& 9.8\times10^{-15} \text{m s$^{-2}$ Hz}^{-1/2} \left(\frac{m_{d,\rm max}}{10^{-9} {\rm g}} \frac{d\rho_d/d\ln m_d}{1.2\times10^{-27} {\rm \, cm}^{-3}} \right)^{1/2}  \left(\frac{M}{10^3 {\rm kg}}\right)^{-1} \left(\frac{A}{10 {\rm m^2}}\right)^{1/2}.\label{eqn:Sadust}
\end{eqnarray}

Figure~\ref{fig:accelerations} shows that the dust acceleration power given by equation~(\ref{eqn:Sadust}) is a subdominant source of acceleration for our fiducial spacecraft specifications.  Only if we increase the maximum mass to $10^{-7} {\rm g}$ does it become comparable to other sources of acceleration, at least in the right panel where $r=30\,$AU.

The accelerations from more massive dust particles can be fit and their contribution to the noise power removed.  The `characteristic strain'  of a single dust event provides an estimate for what dust masses should be detectable and is given by $
h_c^{\rm dust} = {\Delta v}/(2\pi^2 f^2 T_{\rm fit} x)$, where $T_{\rm fit}$ is the period over which other dust collisions contribute less to the acceleration relative to the grain in question. To the extent that the characteristic strain falls above the strain noise, which we find is the case for $m_d \sim 10^{-8} ~(10^{-9})$g if $T_{\rm fit}= 10^4 ~(10^6)$s, the acceleration from the collisions with the dust grain can be fit for and removed.  With our above assumptions about $d \rho_d/d\ln m_d$, grains of mass $m_d$ strike the spacecraft every $\sim 4\times10^6 {\rm~s~} (m_d/10^{-9}{\rm g})^{-1} (A_{\rm eff}/ 10 {\rm m}^2)$, suggesting that dust collisions with $\gtrsim 10^{-9}$g grains can be cleaned.  
This indicates that this source of anomalous accelerations would be removable for dust grain masses that could lead to significant acceleration noise relative to other acceleration sources. 

\subsubsection{Spacecraft charging}
\label{ss:Bfield}

A spacecraft will build up charge as it flows through the interplanetary plasma.  The Lorentz force of the interplanetary magnetic field will then impart accelerations on the spacecraft.  The maximum possible charge $Z_{\rm max}$ is roughly the charge that can repel solar wind protons from striking the spacecraft $1/2 m_p v_{\rm sw}^2 = Z_{\rm max} e^2/R_{\rm sc}$, where $e$ is the electron charge, $m_p$ the proton mass, $v_{\rm sw}$ the velocity of the solar wind, and $R_{\rm sc}$ the characteristic size of the spacecraft.  We find for a velocity characteristic of the solar wind of $v_{\rm sw} = 500\;$km~s$^{-1}$, this results in a maximum voltage of $V \sim  Z_{\rm max} e/R_{\rm sc} = 1300~$V.  

The magnitude of both the homogeneous and inhomogeneous magnetic fields in the outer Solar System are on the order of a $\mu$Gauss, falling off by only a factor of two from 1\;AU to 20\;AU \citep{SMITH1989159}.  The magnetic field power spectrum is also found to maintain a similar spectrum with distance from the Sun.  This near constancy supports our approach of using the magnetic field power spectrum measured by the Voyager~1 spacecraft at $6.1-8.9~$AU to compute the Lorentz force on the spacecraft at all the solar radii considered.  This acceleration power for the maximum spacecraft charge is shown in Figure~\ref{fig:accelerations}, assuming an effective spacecraft extent of $R_{\rm sc} = 2\;$m to calculate $Z_{\rm max}$ and then the acceleration power along an arm is given by $\sqrt{S_{a}} = Z_{\rm max} v_{\rm sw} \sqrt{P_{B}(f, r)}/(\sqrt{3} M)$, assuming isotropic $B$-field perturbations.  
However, this is likely an overestimate as interplanetary spacecraft are designed to have voltages that are likely to be closer to a tenth of our estimate for the maximum \citep{ba3eaa77-fc94-3442-be38-0f8c614f5d3e}.

\subsubsection{Gravity from asteroids and larger bodies}

The gravitational attraction of asteroids will be another source of accelerations.  Using the JPL Small-Body Database, \citet{2021PhRvD.103j3017F} computed the acceleration power from asteroids for spacecraft at $30$AU from the Sun, finding an acceleration power that is absolutely negligible ($\sim 10^{-18} \text{m s$^{-2}$ Hz}^{-1/2}$ at $10^{-7}$Hz and even smaller values at higher frequencies).  However, this catalog of asteroids is wildly incomplete at $30\;$AU.  They also considered a spacecraft at $1~$AU where catalogs are more complete, and found $10^{-12}\text{m s$^{-2}$ Hz}^{-1/2}$ at $< 2\times 10^{-7}$Hz, dropping to $10^{-14}\text{m s$^{-2}$ Hz}^{-1/2}$ at $4\times 10^{-7}$Hz.  Given that even these $1~$AU values are subdominant when compared to the other outer Solar System acceleration sources we compute, asteroids are unlikely to be an important acceleration source.

Displacements owing to gravitational pulls from the Sun and planets should be correctable, as the spacecraft ranging system would measure spacecraft separations to centimeter precision \citep{2023ApJ...947L..23B}. Long-term trends from these pulls on the $\gtrsim$ year orbital timescales of these bodies can be fit for and removed from contaminating the $f\gg $yr$^{-1}$ frequencies of interest.  

\subsubsection{Off-gassing and other spacecraft emissions}
\label{ss:offgassing}

Because our spacecraft would be in such a low acceleration environment, a concern is that off-gassing either from thrusters or from other components in the spacecraft could drive a substantial acceleration. Since external accelerations are $\sqrt{f S_a} \sim 10^{-14}$ m ~s$^{-2}$ for the wave periods of interest, 10~picoNewton of thrust over these periods could result in increased noise.

Thrusters are necessary to dissipate angular momentum on the reaction wheels on week to month times, 
but for the science operation would be generally turned off.  They can be turned on to dissipate angular momentum during a science run, and then their delta-function-like acceleration profile can be fit and removed just like for dust grain collisions.  When turned off, micro-Newton thrusters can only leak at a part in $10^5$ of their baseline thrust to achieve the 10~pico-Newton specification.  
 Additionally, consider the off-gassing of $\sim 280$ K gas from within the thermally regulated spacecraft.  If the spacecraft emits in a single direction, the acceleration would be $a_{\rm off-gas} \sim \Delta m f v/M$ for $\Delta m$ emitted over time $\sim f^{-1}$, or $a_{\rm off-gas} \sim  10^{-14} (\Delta m/ 10^{-4} {\rm g}) (f/10^{-6} {\rm Hz}) (M/10^3{\rm kg})^{-1} {\rm m~s}^{-2}$, where $ 10^{-14} {\rm m~s}^{-2}$ is similar to what we find for external accelerations at a microHertz. As the acceleration power from external sources decreases to higher frequencies, shorter-duration off-gassing events may be even more problematic. Ventilation and thermal control systems that are designed to off-gas perpendicular to the arms and maintain a stable environment over the gravitational wave periods of interest would likely be required to control off-gassing sufficiently to achieve the gravitational wave science.


\section{Strain power sensitivity}
\label{sec:strainpower}

\begin{figure}[htbp]
\centering
\includegraphics[width=1.\textwidth]{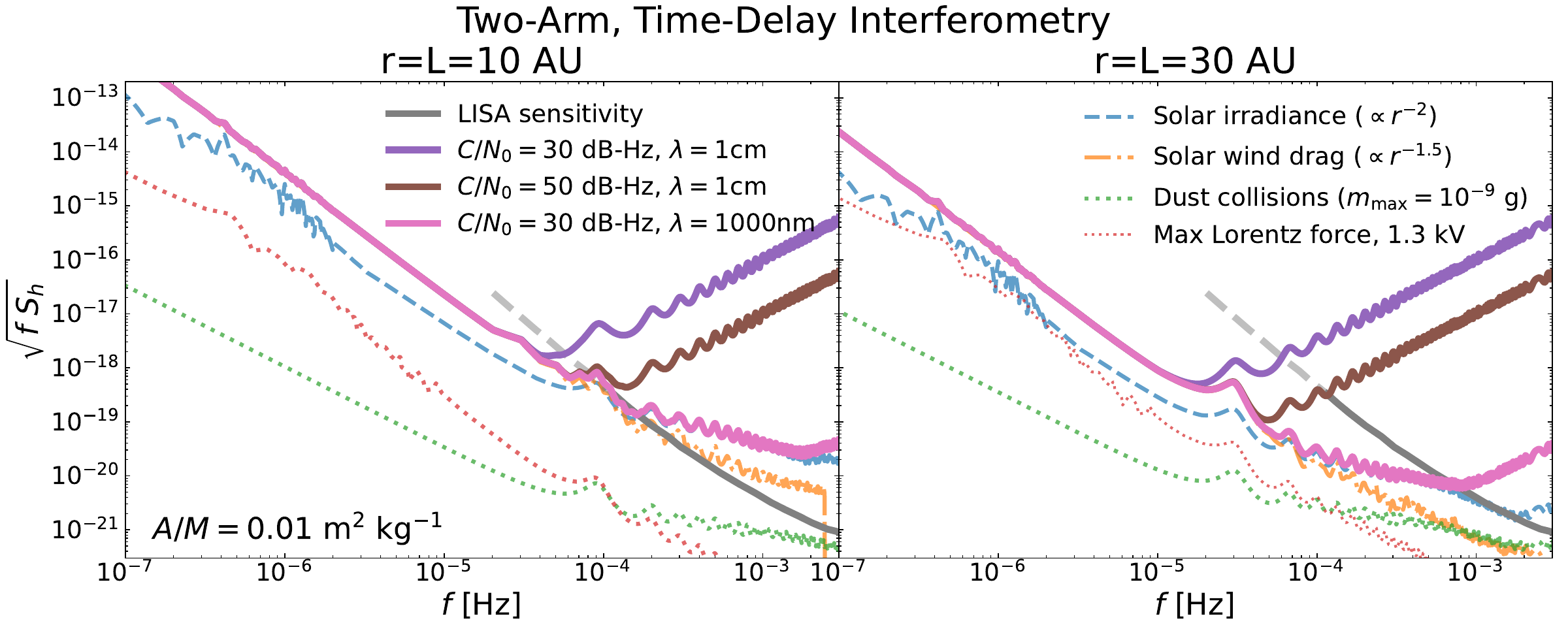}
\caption{Gravitational wave strain sensitivities for the \textbf{first design architecture: the two-arm time-delay interferometry concept}, with $r=L=10$~AU (left panel) and $r= L = 30$~AU (right panel).  The calculations assume an effective area to mass ratio for the spacecraft of $A_{\rm eff}/M = 0.01~$m$^2$~kg$^{-1}$ and a geometric suppression of acceleration noise equal to $0.5$, as would occur if each interferometer arm were a side of an equilateral triangle with the other two sides the Sun-spacecraft distances.   The curves show the contributions from accelerations owing to solar irradiance variations, solar wind drag, the maximum possible Lorentz force from spacecraft charging, and dust collisions assuming $m_{d,\rm max} =10^{-9}$g and $A_{\rm eff} = 10\;$m$^2$.  Also shown by the thick curves are the total strain errors for radio and laser transmissions with different carrier-to-noise ratios ($C/N_0$), where we have excluded the smaller contributions to the accelerations from dust collisions and spacecraft charging.  A requirement of the LISA mission is to achieve the sensitivity shown by the gray solid curve for $f >10^{-4}$Hz, with the goal of being sensitive to  $f > 2\times10^{-5}$Hz as illustrated by the dashed extension.}
\label{fig:strainsensitivity}
\end{figure}

The carrier-to-noise ratio discussed in \S~\ref{ss:shot} and the acceleration noise in \S~\ref{ss:acc} allow us to estimate the gravitational wave strain sensitivity.  Namely, the radiometer noise power, $S_n$, and acceleration power, $S_a$, can be related to the interferometer's sensitivity to the polarization- and sky-averaged gravitational wave strain power at a specific frequency (e.g. \citealt{larson01,2019CQGra..36j5011R}).  The gravitational wave instrument's noise when performing the average that generates this strain power is
\begin{equation}
S_h(f) =  \frac{2}{L^2 R(f)} \left(\frac{ 2{\cal A}^2 ~ S_n}{(2\pi f)^2}  + \frac{4  \big[1 + \cos^2(2\pi f L/c) \big] S_a}{(2\pi f)^4} \right),
\label{eqn:Sh}
\end{equation}
where the factor of $2{\cal A}^2$ is because there are two spacecraft in an arm with independent radiometer noise and ${\cal A}$ encapsulates the increase in error from fitting out plasma dispersion (and is only different from unity in the radio).  The factor $4  \big[1 + \cos^2(2\pi f L/c)\big]$ is because the acceleration of a spacecraft contributes to a displacement that coherently impacts the phase both in the incoming and outgoing directions. The function $R(f)$ is a transfer function that transforms the single-arm noise in the long-wavelength limit (and in which the gravitational wave is propagating orthogonally) to the noise power seen when averaging the instrument's strain response over all angles.  We will first consider the instrument to be a two-arm (time-delay) interferometer and later consider one-arm configurations.   The overall factor of two out front of equation~(\ref{eqn:Sh}) follows the conventions for time-delay interferometry, where the phase observable is a difference between the two independent arms, although we will define $R(f)$ for one-arm configurations to cancel this factor of two so that this equation still applies.  We use the analytic form for $R(f)$ calculated in \citet{larson01}, and for our time-delay interferometer configurations, we assume an angle between the arms of $90^\circ$.\footnote{$R(f) \rightarrow 4/5 \sin^2(\gamma)$ at low frequencies, where $\gamma$ is the angle between arms that we will take to be $90^\circ$.  This differs from the limit $R(f) \rightarrow 6/5$ that applies for LISA by the factor of $2\sin^2(60^\circ)/\sin^2(\gamma)$ owing to LISA having two independent channels at low frequencies and $60^\circ$ orientations for the arms \citep{2019CQGra..36j5011R}.  Equation~(\ref{eqn:Sh}) assumes the noise and accelerations are independent between spacecraft: This will not be true for the acceleration noise on the intermediate spacecraft that joins both arms in the time-delayed inteferometry configuration, as the radial accelerations from solar irradiance and the solar wind will project onto both interferometer arms.} 

In what follows, we present the sensitivity for both a two-arm time-delay interferometry setup (\S~\ref{ss:twoarmforecasts}) as well as a setup with a single arm and atomic clock (\S~\ref{ss:onearm}), and lastly consider the `Doppler tracking' architecture in which the single arm is anchored to Earth (\S~\ref{sec:doppler}).

\subsection{Two-arm time-delay interferometry forecasts}
\label{ss:twoarmforecasts}

The left and right panels in Figure~\ref{fig:strainsensitivity} show the gravitational wave strain sensitivities calculated from equation~(\ref{eqn:Sh}) for the proposed concept at respective solar distances of $r=10$~AU and $r=30$~AU, assuming the same arm lengths as the solar distance, e.g. $L=r$.  The calculations assume a geometric suppression of the radial accelerations from the Sun's radiation and drag forces by $0.5$, as would occur if each interferometer arm were a side of a distinct equilateral triangle with the other two sides the Sun-spacecraft distances.  If instead the configuration were an isosceles triangle with two satellites at distance $r$ from the Sun, the suppression factor would be $L/2r$. These calculations additionally assume an effective area-to-mass ratio for the spacecraft of $A_{\rm eff}/M = 0.01~$m$^2$~kg$^{-1}$, which for example could be achieved with a spacecraft with $A_{\rm eff} = 10$m$^2$ and mass of $M=1000~$kg (perhaps more applicable for the radio dish case) or $A_{\rm eff} = 3~$m$^2$ and $M=300~$kg (perhaps applicable to the laser case and similar to each LISA spacecraft once the solar panels are removed).

The curves in Figure~\ref{fig:strainsensitivity}  show the contributions to the noise from accelerations owing to solar irradiance variations, the solar wind, the maximum possible Lorentz force from spacecraft charging, and dust collisions assuming $m_{d,\rm max} =10^{-9}$g and $A_{\rm eff} = 10\;$m$^2$.  Also shown by the thick curves are the total strain errors for radio and laser transmissions with different carrier-to-noise ratios ($C/N_0$). Those that are for $\lambda=1\,$cm assume carrier-to-noise ratios ($C/N_0$) of $30~$dB-Hz and $50~$dB-Hz (\S~\ref{ss:shot}) and ${\cal A} =1.5$ (Appendix~\ref{sec:dispersion}).  Also shown is a laser effort with $\lambda = 1\mu$m and $30~$dB-Hz.  The noise is smaller for the case of laser links compared to radio ones at $f \gtrsim 5\times10^{-5}\;$Hz, whereas at lower frequencies the sensitivities are similar as acceleration noise dominates. A major goal of the proposed concepts would be to fill in the portion of the gravitational wave spectrum not probed by LISA. The LISA mission (which has three arms and, hence, more interferometric observables than our one-arm concepts) is required to achieve the sensitivity shown by the gray solid curve for $f >10^{-4}\;$Hz, with the goal for LISA to be sensitive to  $f > 2\times10^{-5}\;$Hz as illustrated by the dashed extension.

\begin{figure}[htbp]
\centering
\includegraphics[width=1.0\textwidth]{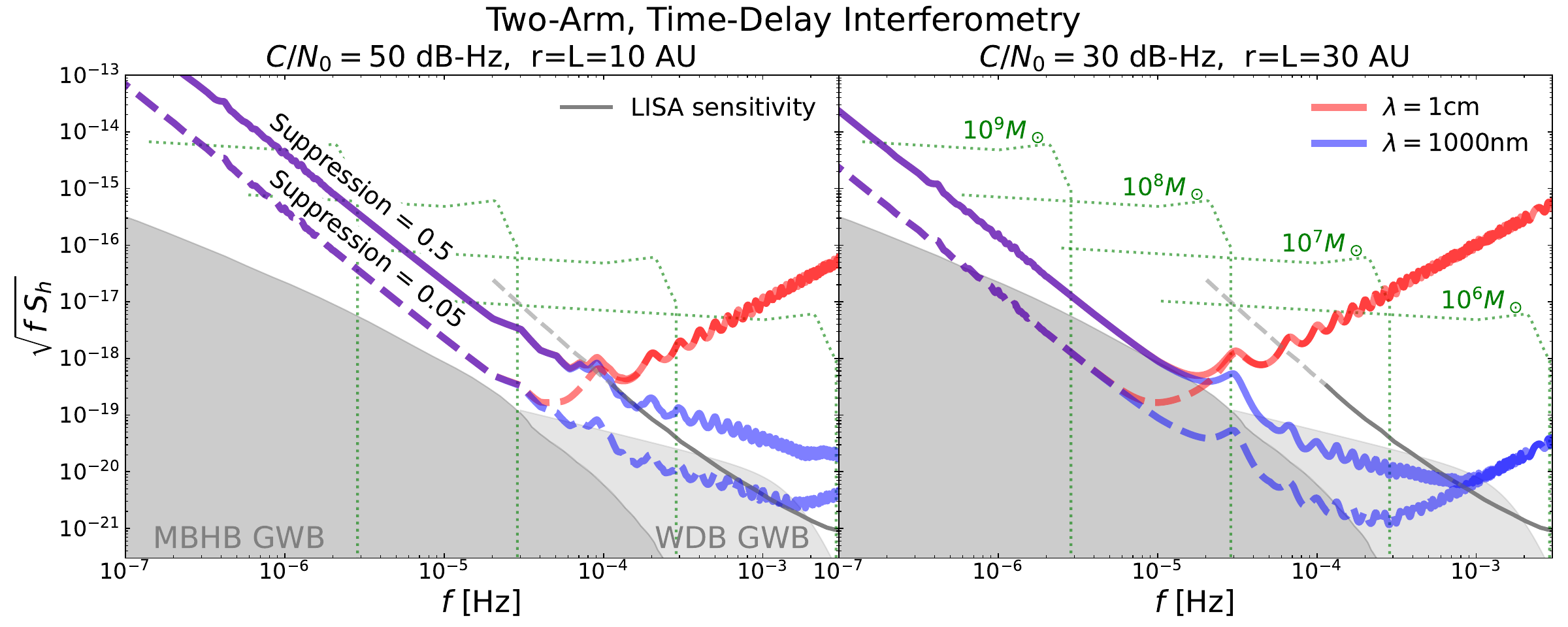}
\caption{Gravitational wave sensitivity for the same time-delay interferometry architectures as shown in Fig.~\ref{fig:strainsensitivity} with $r=L=10\,$AU (left panel) and $r=L=30\,$AU (right panel) and a geometric suppression of acceleration noise equal to $0.5$, plus the same but instead assuming a more aggressive $0.05$ suppression factor.  
Also shown are different astrophysical sources: estimates for the galactic white dwarf binary stochastic gravitational wave background (WDB GWB; \citealt{2017JPhCS.840a2024C}), the massive black hole binary stochastic gravitational wave background (MBHB GWB; \citealt{PhysRevD.84.084004}), and the characteristic strain within $5~$yr of coalescence from equal-mass black hole mergers at $z=3$, where the pre-merger black hole masses are annotated.}
\label{fig:totalsensitivity}
\end{figure}

Figure~\ref{fig:totalsensitivity} shows how these sensitivity projections compare to the gravitational wave signals from astrophysical sources, where we have taken $50\;$dB-Hz for the $r = L = 10\;$AU case (left panel) and $30\;$dB-Hz for the $r=L= 30\;$AU one (right panel).  The green curves show the characteristic strain of equal-mass supermassive black hole mergers at $z=3$ with the initial mass of each black hole annotated and showing the five years before merger \citep{2007CQGra..24S.689A}.  The characteristic strain is defined such that the integral over $\log f$ of the ratio of the characteristic strain to the concept's noise power equals the square of the signal-to-noise ratio for detection.

Figure~\ref{fig:totalsensitivity} also shows the estimate for the Galactic white dwarf binary stochastic gravitational wave background (WDB GWB) from \citet{2017JPhCS.840a2024C}, which we find to be a factor of two lower than the estimate of \citet{2012ApJ...758..131N}.  The stochastic background is defined as the background where the density of sources is too high in each spectral bin over a $\sim 5\;$yr observing period for cleaning to be effective and, hence, this represents essentially an irreducible noise \citep{2017JPhCS.840a2024C}.  Frequencies of $\sim 10^{-4}~$Hz correspond to orbital periods for which gravitational waves in stellar-mass systems are not able to drive coalescence in $\sim 1\;$Gyr.  Since this is approaching the maximum age of stellar systems, at lower frequencies (where the coalescence time is even longer) the gravitational wave background from stellar binaries will be shaped by the initial distribution of orbital properties and not just the limit where this distribution is set by gravitational radiation as these curves assume. 

Below $10^{-4}$ the unresolvable massive black hole binary stochastic gravitational wave background (MBHB GWB) at the centers of galaxies likely exceeds the background from stellar binaries.  The `most likely' estimate for the MBHB GWB from \citet{PhysRevD.84.084004} is shown by the darker gray solid -- and we find the total background in this estimate is in agreement with pulsar timing array observations at $f=3\times10^{-8}\;$Hz \citep[e.g.][]{2023ApJ...951L...6R}.  \citet{PhysRevD.84.084004} also provided `maximum' and `minimum' bounds on the MBHB GWB, with each bound shifting the amplitude of $\sqrt{S_h}$ by a factor of $\sim 5$ relative to the `most likely' model.  

We can also forecast the number of astrophysical sources to which these concepts would be sensitive, using the fact that the strain sensitivities we are forecasting at $f \lesssim 0.5\times 10^{-4}$Hz are somewhat similar to the $\mu$-Ares concept \citep{2021ExA....51.1333S}, particularly for our $r=L=30~$AU configuration. Over a ten year period, $\mu$-Ares forecasts detecting ${\cal O}(1000)$ inspiralling supermassive black holes, finding ${\cal O}(100)$ black hole binaries in the Milky Way, and observing all stars and compact objects that would merge with Sag A$^*$ in $10^6-10^8$ year.  Our most sensitive designs would be capable of similar returns.

Figure~\ref{fig:totalsensitivity} also shows curves for a more aggressive suppression factor of $0.05$.  
Such a suppression would likely require monitoring and then correcting the radiation from the Sun and the solar plasma with additional onboard instrumentation.  In fact, the Lagrange concept, proposed as a potentially cheaper replacement for LISA that reduces the cost by abandoning drag-free control, found that such instrumentation could potentially produce a factor of $100$ suppression beyond the factor of $\sim 10$ they aimed to achieve geometrically \citep{McKenzie2011}. 

\begin{figure}[htbp]
\centering
\includegraphics[width=1.0\textwidth]{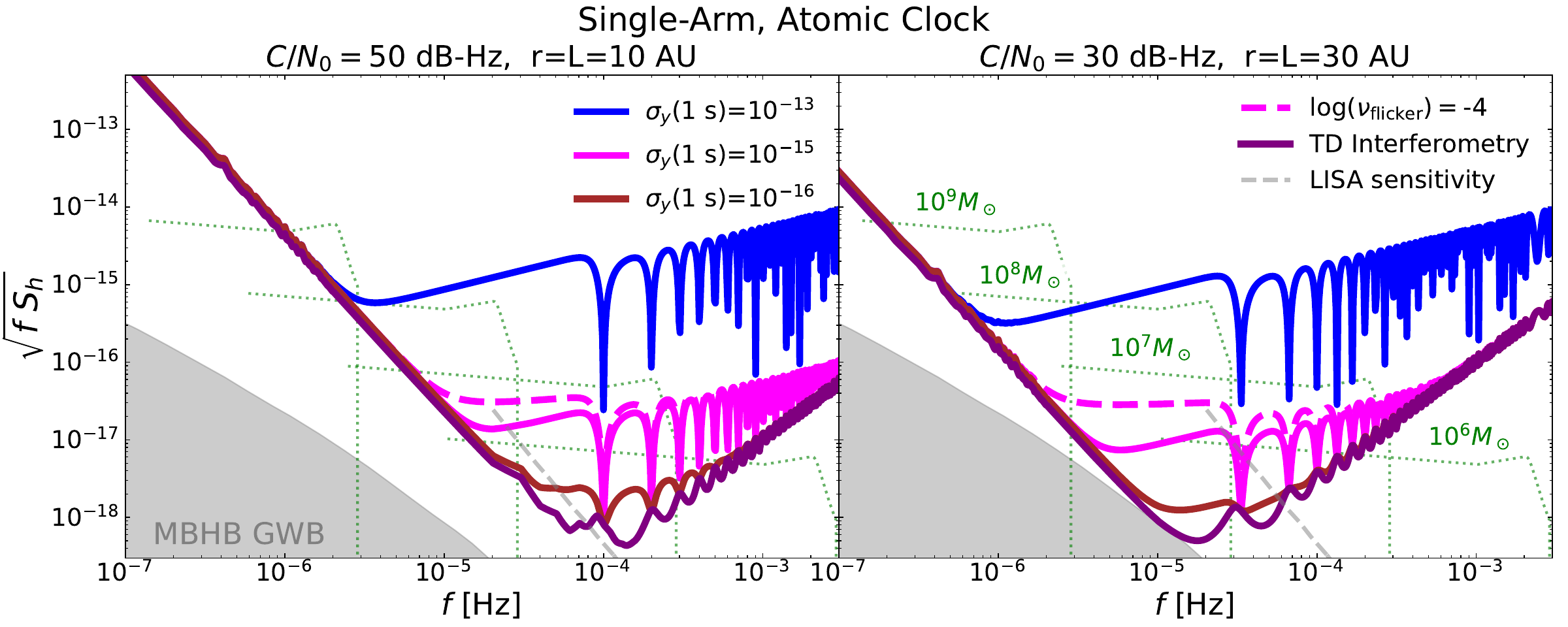}
\caption{Similar to Figure~\ref{fig:totalsensitivity} except showing the gravitational wave strain sensitivities for the \textbf{second design architecture: the single-arm atomic clock concept}.  Additionally, the same $\lambda=1~$cm time-delay interferometry curve as in Figure~\ref{fig:totalsensitivity} is included for comparison.  The curves consider the $\lambda=1\,$cm radio case, although the sensitivity would not be different for the laser case to the extent that clock noise limits the sensitivity.  Allan deviations for $\tau=1\,$s of $\sigma_y = 10^{-13}$ are similar to the Deep Space Atomic Clock, whereas the most precise atomic clocks on Earth achieve $\sigma_y\sim10^{-19}$.  The dashed pink curve shows what happens to the $\sigma_y(\tau = 1 {\rm s}) = 10^{-15}$ case if the clock enters the flicker frequency noise regime at $f<10^{-4}$Hz (rather than white frequency noise as otherwise assumed). Phase measurements referenced to a local clock would achieve nearly the sensitivity of time-delay interferometry with two arms if the clock had $\sigma_y\lesssim 10^{-16}$.  All sensitivity curves assume a geometric suppression of accelerations by a factor of $0.5$.}
\label{fig:clocksensitivity}
\end{figure}

\subsection{Single arm with atomic clock forecasts}
\label{ss:onearm}

A simpler mission would use a single arm, but a single-arm design would also require the inclusion of an atomic clock to reach interesting sensitivities.  Figure~\ref{fig:clocksensitivity} considers this one-arm setup for $r=L=10$ and $30$~AU, but note that the phase noise owing to clock noise is independent of the baseline length for $fL \ll 1$. For this calculation, we generalize equation~(\ref{eqn:Sh}) to also include the clock noise given by equation~(\ref{eqn:phiclock}), requiring the replacement $2 {\cal A}^2 S_n \rightarrow 2 {\cal A}^2 S_n + S_{\rm os, tot}$, where $S_{\rm os, tot}$ is given by equation~(\ref{eqn:phiclock}).  This assumes the signal travels along one arm and then is transponded back to the home satellite that has a precise clock.   
The sensitivity curves are for the $\lambda=1~$cm radio case for the specified $C/N_0$, but we note the laser and radio sensitivities are the same to the extent that the sensitivity is limited by clock noise rather than radiometer noise.

The clock noise is comparable to radiometer noise in the curves that assume the most precise clock with $\sigma_y (\tau = 1 \; {\rm s})=10^{-16}$, as can be seen by noting their sensitivity is comparable to the corresponding time-delay interferometry curve. (The transfer function $R(f)$ is modestly different compared to that of the two-arm designs.\footnote{To compute $R(f)$ for one arm, we drop the $T_3$ term in equation~(23) in \citet{larson01} that accounts for the cross-correlation of the gravitational wave signal between two arms, but include all other terms in their calculation of $R(f)$.  This results in an $R(f)$ that is a factor of two larger than might be most natural to define for a single arm, but this extra factor of two cancels the two in front of equation~(\ref{eqn:Sh}) that owes to noise in both arms, allowing this equation to still apply to our one-arm concepts.})   At larger $f$, distinct downward spikes in the noise are present at $2 L f = n$ for integer $n$, due to the clock noise canceling in the noise power.  One can see that the amplitude of these downward spikes decreases with decreasing $\sigma_y$, as clock noise becomes less of a limiting factor.   

For the case of radio transmissions, phase measurements referenced to a local clock would achieve nearly the sensitivity that would be achieved by time-delay interferometry if the clock has $\sigma_y\lesssim 10^{-16}$.   Allan deviations for $\tau=1\,$s of $\sigma_y = 10^{-13}$ are similar to those of the Deep Space Atomic Clock, whereas $\sigma_y=10^{-16}$ has been bested by three orders of magnitude by the most precise Earth-based clocks \citep{greatatomicclocks}. The calculations discussed so far assume $\sigma_y^2 \propto \tau^{-1}$ as applies in the case of white frequency modulation noise, a scaling demonstrated to hold for $\tau <10^5\;$s in the case of the Deep Space Atomic Clock (but sometimes only to $\tau \sim 10^4\;$s for the most precise atomic clocks). When white frequency modulation noise no longer applies, clocks often enter the `flicker frequency modulation' noise regime where $\sigma_y(\tau)$ is constant with $\tau$.  The pink dashed curves in Figure~\ref{fig:clocksensitivity} show that transitioning to this flicker clock noise scaling at $f<10^{-4}$Hz for the case $\sigma_y(\tau=1{\rm\, s})=10^{-15}$ has a relatively modest affect on the sensitivity.  

\subsection{Doppler tracking forecasts}
\label{sec:doppler}

A potentially simpler design uses Earth for one element in the single-arm setup.  This approach has an extensive history and is called Doppler tracking (for a review, see \citealt{dopplerranging}).  Beyond requiring just a single spacecraft, such a setup has several other advantages: 1) the Earth station's non-gravitational accelerations are potentially negligible; 2) more substantial resources are available at the Earth station, allowing kilowatt uplinks, larger collecting areas, and cryogenic cooling; and 3) the Earth station can be equipped with an ultra-precise atomic clock. However, the mechanical distortions of the instruments from gravity and terrestrial temperature cycles are larger on Earth, and an Earth station has to further contend with propagation delays due to the Earth's atmosphere.  For reference, our estimates for a purely space-based time-delay interferometry setup with $r=L=30\,$ AU have sensitivities of $\sqrt{f S_h} \sim  10^{-16}$ for $f=10^{-6}$Hz, which corresponds to detecting variations of $0.3\,$cm.  
It is conceivable that comparable distance control could be achieved from Earth since lunar laser ranging has achieved $< 0.1~$cm measurements of the Earth-Moon distance \citep{2023PASP..135j4503C}.

Previous Doppler tracking, which has been most successfully executed with the Cassini spacecraft, has been limited by mechanical variations in the analog path owing to e.g. thermal expansion or gravitational stresses. For Cassini, this resulted in centimeter-scale noise over relevant frequencies \citep{dopplerranging}. However, mechanical distortions could potentially be further reduced by using a smaller terrestrial radio dish constructed to be more rigid \citep{dopplerranging}, by using an optical laser setup, or by employing a precise atomic clock on the spacecraft.  More details on the latter two are provided below.  

If mechanical noise could be significantly reduced, atmospheric delays are likely to become the limiting factor. The atmospheric delay that is most worrisome is the tropospheric ``wet'' delay from water vapor in the atmosphere because it is frequency independent -- it cannot be removed by observing multiple frequencies.  Although typical wet delays vary by several centimeters over a day towards zenith, much of this delay can be removed by precise atmospheric monitoring, such as with water vapor radiometers. Indeed, such monitoring occurred for the Cassini Doppler tracking system, for which \citet{Woo_2004} estimated its error by differencing the predictions of two identical systems.  They predicted that the delay could be corrected to $\sim 0.01~$ cm over times of hours to months, although their differencing methodology should be taken as a lower bound on the true error, as it does not account for modeling uncertainties.  

\begin{figure}[ht]
\centering
\includegraphics[width=1.\textwidth]{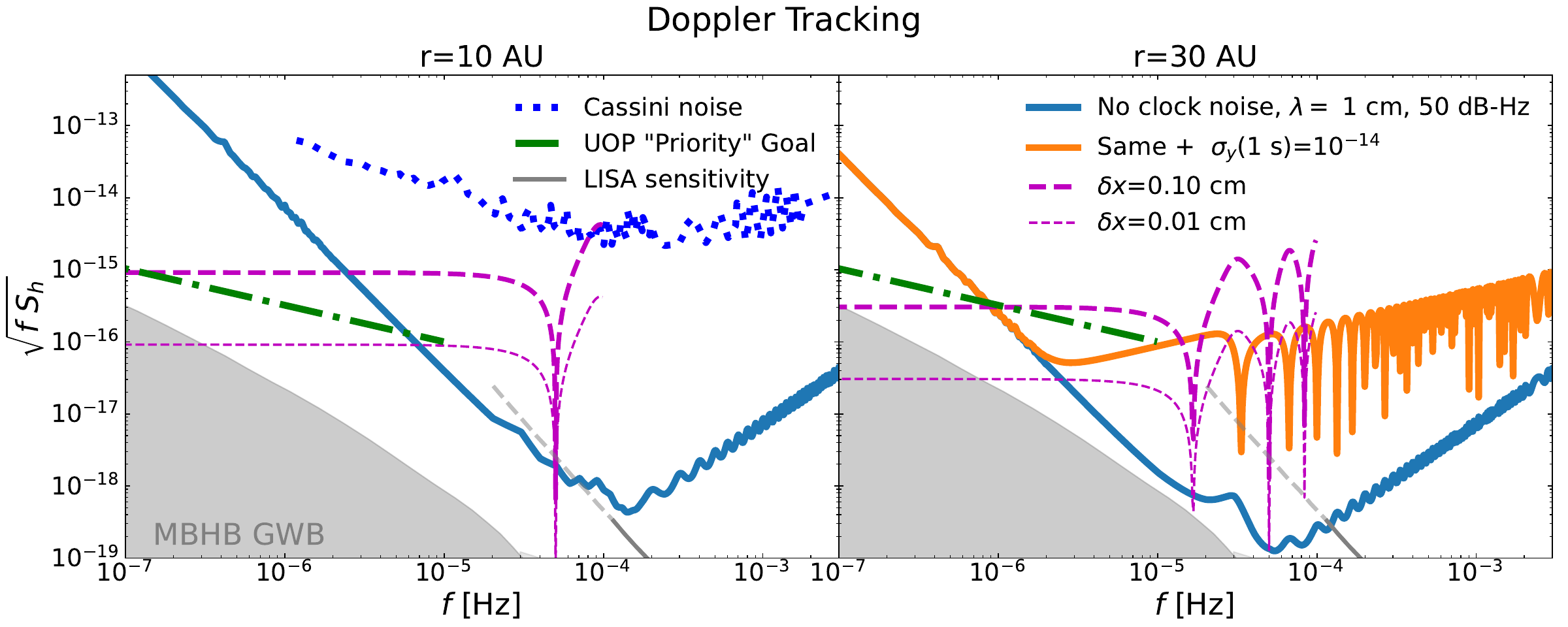}
\caption{Gravitational wave strain sensitivities for the \textbf{third design architecture: the Doppler tracking concept}, sending $\lambda = 1\,$cm radio signals between an Earth station and a $r=10~$AU (left panel) and $r=30~$AU (right panel) spacecraft.  The solid blue curve is the strain sensitivity that an ideal instrument could achieve if its sensitivity were set by spacecraft accelerations and radiometer noise with the specified $C/N_0$.  Mechanical noise and Earth's atmosphere likely limit the sensitivity over this ideal curve.  The magenta dashed curves illustrate the noise power if the atmosphere and mechanical noise could be subtracted to a residual power that had equal variance per $\log f$ and equal to $0.1\,$cm and $0.01\,$cm, numbers motivated in the text. 
The orange Doppler-tracking sensitivity curve in the right panel is the same as the blue one except that it includes clock noise with $\sigma_y(1s)= 10^{-14}$.  If the spacecraft is equipped with such a clock, an observable exists that can remove atmospheric and mechanical delays.  
The green dot-dashed curve shows the $f<10^{-5}$Hz `priority' sensitivity goal for a future mission to Uranus of \citet{2024arXiv240602306Z}. 
The dotted blue points in the left panel are the best constraints using Doppler tracking, achieved with the Cassini mission for which $L\approx 10~$AU  \citep{dopplerranging}.  
}
\label{fig:dopplertracking}
\end{figure}

Figure~\ref{fig:dopplertracking} investigates the potential sensitivity of Doppler tracking of an $L=10~$AU (left panel) and $L=30~$AU (right panel) spacecraft.  The solid blue curve is the strain sensitivity that a Doppler tracking arm could achieve if its sensitivity were set by spacecraft accelerations and the downlink radiometer noise with $C/N_0= 60$ dB-Hz (left panel) and $40$ dB-Hz (right panel) and broadcasting in the Ka band at $\lambda = 1~$cm, with these larger dB-Hz values than in previous plots reflecting that more collecting area could potentially be available for the Earth station.  (The downlink is the limiting step in terms of phase noise.)  Unlike in previous figures, where we assumed some geometric cancellation of the largely radial spacecraft acceleration, the radial nature of Doppler tracking to outer Solar System spacecraft means that nearly all of the acceleration projects onto the Earth-spacecraft chord.  Still, mechanical noise and the atmosphere are likely to limit the sensitivity over this ideal curve.  Indeed, the dotted blue points are the best constraints using Doppler tracking, achieved with the Cassini mission with $L\approx 10~$AU, with the sensitivity likely limited by mechanical noise \citep{dopplerranging}.  Improvements in the Cassini noise by as much as several orders of magnitude would be required to reach the blue solid curve.

 Atmospheric and mechanical delays could also be reduced by using a precise clock on the spacecraft and then doing one-way and two-way ranging (`One-way' uses the clock on the spacecraft as the phase reference that is differenced with a signal sent from Earth. `Two-way' phases the spacecraft to the terrestrial phase and so does not require a precise clock on the spacecraft.  Therefore, this is the mode used by previous Doppler tracking experiments.) Analogous to time-delay interferometry, an observable can be constructed from the phases of the one-way and two-way ranging signals that cancels out any delays that occur at the time of emission and reception by the Earth station and does not remove all of the gravitational wave signal \citep{Armstrong_2021}, namely by differencing the phases in this manner: $\big(\text{two-way}(t)\big)$ - $\big(\text{one-way}(t)\big)$ - $\big(\text{one-way}(t+ L/c)\big)$.  
 The orange curve in Figure~\ref{fig:dopplertracking} is the same as the blue curve but also includes clock noise with $\sigma_y(1s)= 10^{-14}$ -- a value of $\sigma_y(1s)$ that is ten times lower than that of the Deep Space Atomic clock.\footnote{For mechanical and other instrumental delays, the cancellation in the estimator of \citet{Armstrong_2021} occurs to the extent that the one-way and two-way ranging share the same analog path (even during transmit and receive). In our calculations, we have not used the correct transfer function of this observable, which we anticipate will result in a ${\cal O}(1)$ reduction in the sensitivity relative to the single-arm transfer function used for our calculation due to the cancellation of some of the gravitational strain terms in this combination. 
 }  (This type of observable also eliminates acceleration noise at one satellite.  It could motivate a configuration with a more-resourced satellite in the high-acceleration near-Earth environment broadcasting to one in the outer Solar System or, alternatively, the satellite near Earth having a precise accelerometer and the acceleration noise nulled for the outer Solar System satellite.)

The magenta dashed curves shown at $f < 10^{-4}$Hz show the residual noise power if the atmosphere and mechanical noise could be subtracted to a residual power that has an equal variance per $\log f$ (i.e. $1/f$ noise) and equal to $0.1\,$cm and $0.01\,$cm, numbers motivated by the degree to which wet tropospheric delays could be cleaned.   
Although tropospheric delays have a redder spectrum than $1/f$, the spectrum \citet{Woo_2004} found after atmospheric correction was much closer to the $1/f$ form.  
These curves show that for $\lesssim 10^{-5.5}$Hz, achieving $0.01-0.1\,$cm error in mechanical and tropospheric delays would be sufficient to be limited by accelerations in the outer Solar System.\footnote{Our estimates for the noise of a Doppler tracking experiment do not include the error from the Earth's ephemeris. Earth's position needs to be tracked and is uncertain at the $\sim 100\;$m level, with these errors on year- and longer-timescales \citep{2020ApJ...893..112V}.  Ranging can be used to reduce this ephemerides error along the vector to the spacecraft, but this will come at the cost of being less sensitive to gravitational waves on the timescales for which the ephemeris uncertainties need to be fit.  Such fitting will bleed into shorter timescales than Earth's orbital period and possibly set the error for the lowest frequencies considered in our plots of $f\sim 10^{-7}$Hz (as $\sim 100\;$cm residuals will lead to larger errors than accelerations at such frequencies).  A dedicated study is needed to understand the frequencies over which ephemeris errors limit the strain sensitivity.}

There has been some recent interest in Doppler tracking for the $\mu$Hz band in the context of a flagship Uranus mission \citep{2024arXiv240602306Z}.  The green dot-dashed curve in Figure~\ref{fig:dopplertracking} shows the  $f<10^{-5}$Hz `priority' sensitivity goal for this ``Uranus Orbiter and Probe'' (UOP) mission.  This goal assumes an order-of-magnitude improvement of the theoretical Cassini noise, which was an order of magnitude better than what was actually achieved, and extrapolates this noise curve to lower frequencies. Our estimates show that at $f<10^{-6}$Hz, the accelerations on the spacecraft from variations in the solar irradiance and the solar wind must be corrected (likely by equipping the spacecraft with solar irradiance and plasma monitors) to achieve this sensitivity goal.  


As discussed in \citet{2024arXiv240602306Z} for the UOP, Doppler tracking using lasers might also be possible, but such an approach is unlikely to improve the sensitivity with respect to our radio estimates.  Earth's turbulent atmosphere and the low $C/N_0$ downlink make phase locking in the optical impractical (as this requires both adaptive optics and a $C/N_0$ sufficient to achieve phase lock in the millisecond before the atmosphere changes).  Thus, a laser Doppler tracking system would track the beat of laser pulses rather than the carrier phase.  In contrast to timing the carrier phase -- which allows a matched-filter-like approach that results in the timing sensitivity scaling as $L^{-1}$ --, the timing sensitivity for pulsed lasers scales as $L^{-2}$, which is problematic for the envisioned Solar System-scale baselines.  Equation~(\ref{eqn:pulses}) in comparison to equation~(\ref{eqn:hnoiseshot}) shows that at $L=30~$AU the timing noise for pulsed lasers will only be comparable with radio transmissions if a $D_2\sim10\,$m telescope on Earth is deployed along with a $D_1\sim 1\,$m on the spacecraft. Additionally, optical transmissions experience a time-varying delay from the atmosphere that is similar in magnitude to the wet tropospheric delay for radio waves: Near zenith, this delay is a couple of centimeters and can be corrected using atmospheric instrumentation similar to that used for the wet tropospheric delay.  Attempts to correct for this error in the context of lunar laser ranging has resulted in millimeter-scale errors \citep{Marini_1973}.

\section{Instrumental considerations}
\label{sec:control}

At fixed strain noise power due to shot noise, the timing error, displacement error, and square of the angular pointing error are relaxed in proportion to the baseline distance.  Thus, the extremely long arms of the presented concepts may reduce many design tolerances. Of course, long arms do come with the challenge that the transmitted signal strengths are much weaker compared to, e.g., LISA; Section~\ref{ss:shot} argued that the strengths are still sufficient to acquire stable phase locks.  We now discuss other challenges of the laser and radio concepts, as well as challenges with placing spacecraft in the outer Solar System.

\subsection{Laser design}
\label{ss:laserdesign}

 Perhaps the biggest challenge for our laser-based concepts is tuning the laser frequencies to cancel the relative velocities of the spacecraft.  In all the scenarios we considered, the spacecraft would likely have much larger relative velocities compared to LISA, which are kept to $\lesssim 10\,$m~s$^{-1}$.\footnote{The way to avoid large relative velocities is if the outer Solar System spacecraft were in a circular orbit, each separated by $120^\circ$ in orbital phase; \citealt{Folkner2011, 2021ExA....51.1333S}. Achieving such a configuration in the outer Solar System would be challenging.}  Relative velocities are problematic in that they spoil the cancelation of clock errors from time-delay interferometry, as the observed phase difference (eqn.~\ref{eqn:Xt}) now has a strongly time-dependent phase given by $\phi_{\rm het} = 2\pi f_{\rm het}t$, where $f_{\rm het}$ is the part of the Doppler shift to the lasers' emitted frequency that is uncorrected by any frequency tuning.  This time-dependent phase means that, when it is recorded, the clock \recentchange{drift} $\delta t_{\rm os}$ -- or any timing error -- again enters and results in a phase error of
\begin{equation}
\delta \phi_{\rm het} = 2\pi f_{\rm het} \delta t_{\rm os}.
\label{eqn:phasetimeing}
\end{equation}
For LISA, $\delta \phi_{\rm het}$ is large enough to substantially reduce instrument performance.  The LISA design includes an elaborate scheme for correcting this error by superimposing weak pilot tones on top of the primary laser signal. The phase of these pilot tones contains the $\delta t_{\rm os}$ information that allows $\delta \phi_{\rm het}$ to be corrected and essentially eliminated from the phase noise.  The low $C/N_0$ of the inter-spacecraft broadcasts in our concepts may make this strategy of superimposing weaker pilot tones not viable.\footnote{\recentchange{The tracking of the main tone will eliminate some of the low-frequency timing noise for phase tracking the pilot tones, allowing phase locks on the pilot tones with lower $C/N_0$ than the main tone. The viability of using pilot tones with our concepts' weak beams could merit further investigation.}}

\recentchange{Here we investigate whether the elimination of pilot tones entirely could be possible.}  Our concepts' much larger phase error tolerance may allow $\delta \phi_{\rm het}$ to be sufficiently small as to not dominate the error if the spacecraft adjoining both arms is equipped with an atomic clock.  We can write the standard deviation of $\widetilde {\delta \phi}_{\rm het}$ in terms of the Allan deviation of its clock (cf. eqn. \ref{eqn:Sphiclock}):
\begin{eqnarray}
\widetilde{\phi}_{\rm het, rms}  =  \frac{\sqrt{2 \tau  } (2 \pi f_{\rm het})  \sigma_y(\tau)}{(2 \pi f)} =   0.1 ~~\text{rad~Hz}^{-1/2} \left( \frac{\sigma_y(1~ \rm s)}{10^{-13}}  \right)  \left(  \frac{f_{\rm het}}{100 ~\rm MHz} \right)  \left(  \frac{f}{10^{-5}~\rm Hz} \right)^{-1}.
\end{eqnarray}
For reference, the shot noise errors on the phase are $0.03~\text{rad~Hz}^{-1/2}$ for $[C/N_0]_{\rm dB-Hz} = 30\,$dB-Hz (eqn.~\ref{eqn:phirmsshot}).  
We conclude that the heterodyne phase error is manageable with an atomic clock like the Deep Space Atomic Clock, for which $\sigma_y(1~ \rm s) \sim 10^{-13}$ if $f_{\rm het} \sim 10^2$ MHz, an order of magnitude higher than the maximum heterodyne frequency allowed for LISA.  

The simplest orbits would send the spacecraft on radial trajectories with respect to the Sun.  In this case, the relative velocities between the spacecraft could be tens of kilometers per second. A $10\,$km~s$^{-1}$ offset velocity would have to be compensated for by a $\sim 0.3$\AA\ shift of the output wavelength of a $\lambda =10^4$\AA\ laser in order for it to interfere within the heterodyne specification.  Tuneable lasers that can adjust their wavelength even to $>1$\AA\ exist. Additionally, space interferometers like LISA must adjust the laser frequency on month timescales to compensate for this drift and stay under the maximum $f_{\rm het}$ that the system can tolerate \citep{Barke2015thesis, PhysRevD.110.042002}.   For spacecraft on radial trajectories, the rate of change of the frequency from Doppler shifting between a spacecraft and some fixed reference is
\begin{equation}
\frac{d f}{d t} = \frac{G M_\odot}{r^2 \lambda}  = 2\times10^9 \left(\frac{r}{30 \rm AU } \right)^{-2} \left(\frac{\lambda}{1000 \ \text{\AA}} \right)^{-1}~~~~~\text{GHz ~yr$^{-1}$}.
\label{eqn:dnudt}
\end{equation}  Equation~(\ref{eqn:dnudt}) suggests that for radial trajectories and spacecraft at $r=10-30~$AU, this tuning would also have to be done weekly or monthly to maintain a maximum heterodyne frequency of $\sim 100~$MHz.

A second challenge with lasers is controlling the phase noise from laser intensity variations, called `relative intensity noise' (RIN), which scales inversely with received laser power, and the received power is exceptionally small owing to our concepts' long baselines.  The square root of the phase noise power spectrum from RIN is $\widetilde{\phi}^{\rm RIN}_{\rm rms} \approx \sqrt{P_{\rm local}/(2 \eta_{\rm het} P_{\rm rec})}~ r(f_{\rm het}) \times \chi$ , where $P_{\rm local}$ is the power of the local laser that is being recombined with the received beam prior to phase readout, $r(f_{\rm het})$ is the RIN at the heterodyne frequency, and $\chi<1$ expresses how much of this term is canceled by `balanced' detection methods that split the laser beams and combine their phase readout in a manner that, if perfectly performed, eliminates this dominant RIN term \citep{Barke2015, PhysRevApplied.17.024025}.  As RIN phase noise scales in the same manner with $P_{\rm rec}$ as the phase noise from shot noise, it is helpful to take the ratio with the analogous shot noise (eqns.~\ref{dbHzlaser} and \ref{eqn:phirmsshot}):
\begin{equation}
\frac{\widetilde{\phi}^{\rm RIN}_{\rm rms}}{\widetilde{\phi}_{\rm rms}^{\rm shot} } = \sqrt{\frac{P_{\rm local}}{h c/\lambda}} \chi \, r(f_{\rm het}) = 0.7 \, \chi\, \left(\frac{P_{\rm loc}}{10^{-3} ~\text{W}} \frac{\lambda}{1 ~\mu {\rm m}} \right)^{1/2} \left( \frac{r(f_{\rm het})}{10^{-8} ~\text{Hz}^{-1/2}}\right),
\end{equation}
where we have referenced the latter expression to the LISA requirement $r(f_{\rm het}) \approx 10^{-8}\;$Hz$^{-1/2}$ and a $P_{\rm local}$ similar to that of LISA.\footnote{Electronics noise can also add to the phase noise. We further find, using the online interface associated with \citet[{\it Towards a Gravitational Wave Observatory Designer: Sensitivity Limits of Spaceborne Detectors}][]{Barke2015}, that a LISA-like $P_{\rm local}= 2\times 10^{-3}~$W minimizes the combination of shot, RIN and electronics noise for a system with $L=30\;$AU.  Aside from $L$, these calculations assume LISA-like specifications for other system parameters.  At this minimum, the shot noise of the received laser does indeed contribute most of the phase noise rather than RIN and electronics noise (and this interface assumes no balanced detection such that $\chi=1$).  Additionally, for a much larger value of $f_{\rm het}=500~$MHz from $f_{\rm het}=25~$MHz that was used for the calculations just summarized, which increases electronics noise, we find the total noise from all three terms is minimized at $\sim 3\times$ shot noise alone when $P_{\rm local} = 0.03~$Watt.}    Thus, especially for balanced detection that potentially could achieve $\chi <0.1$ \citep{PhysRevApplied.20.014016}, RIN is likely not a limiting factor for the phase tracking system as long as the lasers achieve LISA-like specifications for $r(f_{\rm het})$.

\subsection{Radio design}

There are many aspects that are easier for an experiment that relies on radio transmissions. Requirements on any error that manifests as an apparent satellite displacement are relaxed relative to laser transmissions by the $\sim 10^4$ ratio of optical to radio wavelengths (at least frequencies where the sensitivity is set by radiometer noise).  Because the phase of the signal itself would be fed directly into the phase-lock loop for radio transmissions, in contrast to the laser setup in which the phase of the received laser is beat against a reference laser, spacecraft relative velocities and relative intensity noise are not a concern.  The pointing requirements are also relaxed relative to lasers, as lasers generate more planar wavefronts than radio dishes (which leads to larger phase errors from mispointing). 

The radio design requires large dishes and an analog instrumental path via wires, such that thermal path length variations are likely to be larger than the optical design. 
For metals, the thermal expansion factor for $1\;$m of path length is typically $\sim 10^{-6} - 10^{-5}\,$K$^{-1}$, with the smaller values being for temperatures an order of magnitude below room temperature.  To keep thermal expansion to $10^{-4}\,$cm, as required for these path length changes to be a tenth or so of the radiometer noise (eqn. \ref{eqn:xrms}), the temperature needs to be controlled to just tens of Kelvin. 

\subsection{Outer Solar System considerations}

One challenge to our proposal is that outer Solar System missions are significantly restricted when it comes to downlink data rates, their mass, and the power budget. Here we discuss each.

\begin{description}
\item[downlink data rates] Appendix~\ref{sec:datarates} shows that, due to the low gravitational wave frequencies targeted, even $10\,$kbps hour-long downlinks every several months could be sufficient.  Such downlink rates have been achieved to outer Solar System spacecraft, including New Horizons.

\item[mass and orbits] The New Horizons spacecraft took nearly a decade to reach Pluto at 34 AU despite weighting just $500\;$kg \citep{2008SSRv..140...23F}.  However, with new Block 1B and Block 2 rockets from the Space Launch System, plus advances in third/fourth stage boosters, it will soon be possible to launch several times more massive spacecraft at the same Earth-escape velocity as New Horizons \citep{nasa2020sls} and it is likely that similar specifications will be met by SpaceX and Blue Origin rockets.  These new rockets would allow a spacecraft with the mass of New Horizons to reach Pluto in half the time. As achieving a particular orbit would require substantial fuel to slow down, the most feasible architecture would likely involve spacecraft that are continually drifting farther from the Sun, possibly on unbound trajectories from the Solar System.\footnote{Spacecraft trajectories that do not require achieving precise orbits may be an advantage over other interferometric concepts targeting microHertz gravitational waves, which require precise placement at $120^\circ$ in orbital phase at $1-3~$AU \citep{Folkner2011,2023CQGra..40s5022M, 2021ExA....51.1333S, 2023CQGra..40s5022M}.}

\item[power] Outer Solar System spacecraft must rely on radioisotope power.  If the power output is similar to the radioisotope thermoelectic generator on the New Horizons spacecraft, the total power budget would be $250~$Watt, and only a fraction of this could be dedicated to the spacecraft's science system.  Meeting such a restrictive power budget seems potentially feasible given our result that $\sim 1- 10~$Watt inter-spacecraft transmissions are sufficient.  We further estimate similar power requirements for other systems, such as for compute and for the reaction wheels.  For our designs that rely on an ultra-precise atomic clock, this technology likely would require substantial development.  The Deep Space Atomic Clock requires 47~Watt \citep{Burt2020}, although there is a miniaturized version of this trapped ion clock that requires 6~Watt but has an order of magnitude larger Allan deviation \citep{m2picclock}.  Clocks based on optical frequency combs are being developed by NASA and likely can be smaller and less power-intensive, in addition to having smaller Allan deviations \citep{Tomio_2024}.  Power budgeting for our concepts is aided by removing the requirement of ultra-precise drag-free control -- one of the most power-intensive systems on LISA. 
\end{description}

\subsection{Systematic checks and sky localization considerations}
\label{ss:localization}

For gravitational wave observatories, there are two more important considerations to make when designing the mission - the ability to perform systematic checks, and the ability to localize gravitational wave signals on the sky.  Systematic checks are necessary for being able to identify and differentiate occasional transient signals as being either instrumental in nature (``glitches'') vs. being astrophysical in nature (``bursts'').  And in general, sky localization to any source tends to be poorer for gravitational wave observatories as compared to electromagnetic observatories, since they are effectively all-sky antennas.

Considering terrestrial-based gravitational wave observatories, the duration of the signals they are sensitive to is short, typically on the order of seconds to minutes.  During this time, the antenna pattern of an individual interferometer observatory does not change a significant amount; therefore, multiple observatories placed at very large separations across the Earth and in different orientations are necessary in order to estimate a sky location for any given signal.  Having multiple observatories also benefits their systematic checking capabilities, as localized transient glitches in any one observatory would not appear across multiple, far-separated observatories.  Astrophysical gravitational bursts would, however, appear correlated across all observatories.  This helps give terrestrial observatories their glitch vetoing capabilities.

Considering the LISA mission, the duration of the signals they will be sensitive to is much longer, ranging between hours to years.  During this time, the antenna pattern of the observatory will change as the constellation of satellites tumbles and orbits the Sun.  The changing antenna pattern means that once an individual source has been identified, its localization will improve over time.  The shorter the duration of the signal (e.g. for astrophysical bursts), the poorer the sky localization will be.

It is also anticipated that glitch vetoing will be partially enabled by having three time-delay interferometry data channels, given the three-arm design.  Isolated transient instrumental noise signals occurring on one of the three spacecraft will propagate through the data channels differently than a common astrophysical burst that hit all three spacecraft. Therefore, in considering the three architectures presented in this work, these two factors may favor a multi-arm design.  This fits with the two (or more) arm, time-delay interferometry architecture, or multiple concurrent instances of single-arm missions, enabled either by atomic clocks or through Doppler tracking.  

The angular resolution of a single arm with length $L$ we anticipate would be $\delta \theta \sim \lambda_{\rm GW}/(L~ {\rm SNR})$, with no rotational information around each arm and a $\pm \theta$ degeneracy, where SNR is the signal-to-noise ratio of the gravitational wave event, and this assumes that for $\lambda_{\rm GW} \lesssim L$ the source drifts appreciably in frequency in order to select the correct `interference fringe' (see Appendix~\ref{ap:angres} for additional details).  Since our concepts target extremely long wavelengths of $\lambda_{\rm GW} = 2000~ (1 \mu \text{Hz}/f)\;$AU, localizations $\delta \theta \lesssim 10^\circ$ that would be most useful for electromagnetic follow-up will only be possible for bright sources that appear at the highest frequencies these concepts are potentially sensitive (likely $f \gtrsim 10^{-4}\;$Hz).  Furthermore, analogous to the LISA spacecraft, the spacecraft would likely drift by many astronomical units over year timescales, reaching different points in the phase pattern of a long-term gravitational wave source, which could be further used for localization and to isolate gravitational wave signals from systematics.


\section{Conclusions}

This paper discussed the feasibility of detecting microhertz gravitational waves using outer Solar System spacecraft. This waveband probes the merger of supermassive black holes as well as a host of other gravitational wave phenomena. Taking advantage of the low acceleration environment beyond $10$~AU, such a system could avoid the substantial technological development required for sufficient drag-free control at lower frequencies of $f<2\times10^{-5}$Hz than the LISA mission targets. For solar distances as well as inter-spacecraft separations of $\geq 10$~AU, we showed that the various interplanetary acceleration sources are small enough that even reaching a sensitivity where the noise is set by the stochastic gravitational wave background from massive black holes and white dwarf binaries appears achievable.  We showed that such an acceleration-limited system would be easily able to detect the mergers of supermassive black holes at all likely redshifts.  

We investigated systems that lock onto and time the phase from both laser and radio transmissions between the spacecraft.  For both, we argued that even for $30\;$AU separations, transmission powers of $\sim 10~$Watt and reasonable mirror/dish sizes, stable phase locks may be achievable.  We found that for the laser concepts this would require reduced frequency noise compared to the allowance for the LISA lasers, although possibly in line with the frequency noise achieved by the lasers in the GRACE-FO mission.  Additionally, despite the much longer wavelengths of radio compared to laser transmissions, we showed that the sensitivity is likely to still be set by acceleration noise at $\lesssim 3\times 10^{-4}~$Hz and, hence, independent of the wavelength of the transmissions. A system that uses the radio significantly reduces many design tolerances, such as those regarding pointing, transmission intensity variations, and spacecraft relative velocities. For radio implementations, interplanetary plasma contributes phase noise that can be effectively eliminated by transmitting at two wavelengths, with only a modest (factor of $\sim 1.5$) reduction in strain sensitivity, and then only when limited by radiometer noise. 

This paper considered three possible architectures.  The first was a two-arm (three-spacecraft) configuration that allows time-delay interferometry.  This configuration was also the most sensitive without substantial improvement in space-certified atomic clocks.  We considered configurations where the arms are at solar distances of $10$ and $30~$AU, and where the arm lengths were the same as the solar distance.  Both configurations were able to detect merging $10^7-10^{10}M_\odot$ black holes out to substantial redshifts, and the $30~$AU case was sensitive to middle-of-the-road predictions for the stochastic gravitational background at $\sim 10^{-5}$Hz.  The sensitivity can be further improved by correcting for accelerations by monitoring solar irradiance variations and the solar wind or, alternatively, with onboard acceleration control.  

The second architecture we considered involved just a single arm.  As a single arm cannot do interferometry to essentially eliminate clock noise, a single-arm design must incorporate a precise atomic clock on at least one spacecraft.  We showed that the single-arm architecture equipped with a clock similar to the Deep Space Atomic Clock -- a clock scoped for future interplanetary missions -- could be sensitive to the characteristic strains of $10^8M_\odot$ and $10^9M_\odot$ supermassive black hole mergers, one of the most exciting signals anticipated in the microhertz waveband.  Three orders of magnitude improvements in the timing precision over the Deep Space Atomic Clock, still far from the precision of the most precise terrestrial clocks, could achieve a sensitivity similar to the interferometric configuration when comparing at the same $L$.

The single-arm design becomes Doppler tracking of outer Solar System spacecraft when one of the nodes is located on Earth.  Doppler tracking using outer Solar System spacecraft has a rich history \citep{dopplertracking}. Doppler tracking is the final architecture that we considered.  We discussed the atmospheric and ground station delay requirements for Doppler tracking to reach interesting sensitivity benchmarks.  We showed that for gravitational waves with $f\lesssim 1\mu$Hz, spacecraft accelerations must be corrected with onboard instrumentation for Doppler tracking to achieve the sensitivity goals of \citet{2024arXiv240602306Z}, envisioned in the context of a future Uranus probe.  
We also investigated the sensitivity of Doppler tracking if the spacecraft could be equipped with an atomic clock, allowing for a time-delay observable that nulls out atmospheric and some mechanical delays.

Placing spacecraft in the outer Solar System puts severe limits on the downlink rates, mass requirements, and power considerations.  We showed that the achievable downlink rates should be sufficient because of the low frequencies of the targeted gravitational waves.  We also argued that several tens of Watts of power could conceivably power the science systems on the spacecraft, within the realm of what can be supplied by a radioisotope thermoelectric generator.  The mass of each spacecraft would likely have to be under $10^3$kg in order to be launched to tens of astronomical units in $5-10~$yr.  The spacecraft requirement that we identified as potentially concerning to operate without onboard acceleration monitoring is the severe restrictions on off-gassing (\S~\ref{ss:offgassing}).  

Although the discussion in this paper focused on concepts without acceleration control and $\mu$Hz gravitational waves, some of our results could also apply to an outer Solar System concept that includes an ultra-precise accelerometer or that targets higher frequencies.  The extremely long arms of our hypothetical concepts mean that the accelerometer would not need to be as precise as in designs with shorter arms to reach the same sensitivity. Furthermore, the stable thermal and acceleration environment of the outer Solar System may facilitate acceleration control over the long periods of our targeted gravitational waves.  Laser locks over $\sim 10~$AU arms could allow better sky localizations at $f \lesssim 10^{-4}$Hz than more LISA-like concepts.\\

\section*{Acknowledgments}
The authors thank Eric Agol, Miguel Morales, and Joseph Lazio for useful discussions.  We thank Jacob Slutsky, Shimon Kolkowitz, and Lennart Wissel for feedback on an earlier version of the manuscript.  This work is supported by NASA/NIAC Phase I award 80NSSC24K0644.  C.M. would like to acknowledge support by NASA under award number 80GSFC24M0006.

\textit{Materials Availability.}  The code for the calculations presented in this work can be found at \url{https://github.com/astromcquinn/GWwithDragFree.git}. 

\textit{Software.}  This work made use of the following software packages: \texttt{Jupyter } \citep{2007CSE.....9c..21P, kluyver2016jupyter}, \texttt{matplotlib } \citep{Hunter:2007}, \texttt{numpy } \citep{numpy}, and \texttt{python } \citep{python}.  Software citation information aggregated using \texttt{\href{https://www.tomwagg.com/software-citation-station/}{The Software Citation Station}} \citep{software-citation-station-paper}. 


\bibliographystyle{mnras}
\bibliography{References}

\begin{thebibliography}{}
\makeatletter
\relax
\def\mn@urlcharsother{\let\do\@makeother \do\$\do\&\do\#\do\^\do\_\do\%\do\~}
\def\mn@doi{\begingroup\mn@urlcharsother \@ifnextchar [ {\mn@doi@} {\mn@doi@[]}}
\def\mn@doi@[#1]#2{\def\@tempa{#1}\ifx\@tempa\@empty \href {http://dx.doi.org/#2} {doi:#2}\else \href {http://dx.doi.org/#2} {#1}\fi \endgroup}
\def\mn@eprint#1#2{\mn@eprint@#1:#2::\@nil}
\def\mn@eprint@arXiv#1{\href {http://arxiv.org/abs/#1} {{\tt arXiv:#1}}}
\def\mn@eprint@dblp#1{\href {http://dblp.uni-trier.de/rec/bibtex/#1.xml} {dblp:#1}}
\def\mn@eprint@#1:#2:#3:#4\@nil{\def\@tempa {#1}\def\@tempb {#2}\def\@tempc {#3}\ifx \@tempc \@empty \let \@tempc \@tempb \let \@tempb \@tempa \fi \ifx \@tempb \@empty \def\@tempb {arXiv}\fi \@ifundefined {mn@eprint@\@tempb}{\@tempb:\@tempc}{\expandafter \expandafter \csname mn@eprint@\@tempb\endcsname \expandafter{\@tempc}}}

\bibitem[\protect\citeauthoryear{{Abbott} et~al.}{{Abbott} et~al.}{2016}]{PhysRevLett.116.061102}
{Abbott} B.~P.,  et~al., 2016, \mn@doi [Phys. Rev. Lett.] {10.1103/PhysRevLett.116.061102}, 116, 061102

\bibitem[\protect\citeauthoryear{{Abbott} et~al.,}{{Abbott} et~al.}{2017}]{2017PhRvL.119p1101A}
{Abbott} B.~P.,  et~al., 2017, \mn@doi [\prl] {10.1103/PhysRevLett.119.161101}, \href {https://ui.adsabs.harvard.edu/abs/2017PhRvL.119p1101A} {119, 161101}

\bibitem[\protect\citeauthoryear{{Abbott} et~al.,}{{Abbott} et~al.}{2023}]{2023PhRvX..13d1039A}
{Abbott} R.,  et~al., 2023, \mn@doi [Physical Review X] {10.1103/PhysRevX.13.041039}, \href {https://ui.adsabs.harvard.edu/abs/2023PhRvX..13d1039A} {13, 041039}

\bibitem[\protect\citeauthoryear{{Aeppli}, {Kim}, {Warfield}, {Safronova}  \& {Ye}}{{Aeppli} et~al.}{2024}]{2024PhRvL.133b3401A}
{Aeppli} A.,  {Kim} K.,  {Warfield} W.,  {Safronova} M.~S.,   {Ye} J.,  2024, \mn@doi [\prl] {10.1103/PhysRevLett.133.023401}, \href {https://ui.adsabs.harvard.edu/abs/2024PhRvL.133b3401A} {133, 023401}

\bibitem[\protect\citeauthoryear{Agazie et~al.,}{Agazie et~al.}{2023}]{Agazie_2023}
Agazie G.,  et~al., 2023, \mn@doi [The Astrophysical Journal Letters] {10.3847/2041-8213/acdac6}, 951, L8

\bibitem[\protect\citeauthoryear{{Ajith} et~al.,}{{Ajith} et~al.}{2007}]{2007CQGra..24S.689A}
{Ajith} P.,  et~al., 2007, \mn@doi [Classical and Quantum Gravity] {10.1088/0264-9381/24/19/S31}, \href {https://ui.adsabs.harvard.edu/abs/2007CQGra..24S.689A} {24, S689}

\bibitem[\protect\citeauthoryear{{Arca Sedda} et~al.,}{{Arca Sedda} et~al.}{2020}]{2020CQGra..37u5011A}
{Arca Sedda} M.,  et~al., 2020, \mn@doi [Classical and Quantum Gravity] {10.1088/1361-6382/abb5c1}, \href {https://ui.adsabs.harvard.edu/abs/2020CQGra..37u5011A} {37, 215011}

\bibitem[\protect\citeauthoryear{Armstrong}{Armstrong}{2006a}]{dopplertracking}
Armstrong J.~W.,  2006a, \mn@doi [Living Reviews in Relativity] {10.12942/lrr-2006-1}, 9, 1

\bibitem[\protect\citeauthoryear{Armstrong}{Armstrong}{2006b}]{dopplerranging}
Armstrong J.~W.,  2006b, \mn@doi [Living Reviews in Relativity] {10.12942/lrr-2006-1}, 9, 1

\bibitem[\protect\citeauthoryear{Armstrong}{Armstrong}{2021}]{Armstrong_2021}
Armstrong J.~W.,  2021, Gravitational Wave Search With the Clock Mission, \url {https://ntrs.nasa.gov/citations/20210005049}

\bibitem[\protect\citeauthoryear{Ascheid \& Meyr}{Ascheid \& Meyr}{1982}]{cycleslips}
Ascheid G.,  Meyr H.,  1982, \mn@doi [IEEE Transactions on Communications] {10.1109/TCOM.1982.1095423}, 30, 2228

\bibitem[\protect\citeauthoryear{Bachman et~al.,}{Bachman et~al.}{2017}]{Bachman_2017}
Bachman B.,  et~al., 2017, \mn@doi [Journal of Physics: Conference Series] {10.1088/1742-6596/840/1/012011}, 840, 012011

\bibitem[\protect\citeauthoryear{{Bai}, {Chen}  \& {Korwar}}{{Bai} et~al.}{2023}]{2023JHEP...12..194B}
{Bai} Y.,  {Chen} T.-K.,   {Korwar} M.,  2023, \mn@doi [Journal of High Energy Physics] {10.1007/JHEP12(2023)194}, \href {https://ui.adsabs.harvard.edu/abs/2023JHEP...12..194B} {2023, 194}

\bibitem[\protect\citeauthoryear{Barke}{Barke}{2015}]{Barke2015thesis}
Barke S.,  2015, PhD thesis, Gottfried Wilhelm Leibniz Universität Hannover, Hannover, Germany, \url {https://myaidrive.com/file-2T5njY8LayKGJL0BMyABpAhk}

\bibitem[\protect\citeauthoryear{{Barke} et~al.,}{{Barke} et~al.}{2014}]{Barke2014LISAMS}
{Barke} S.,  et~al., 2014. \url {https://api.semanticscholar.org/CorpusID:106763425}

\bibitem[\protect\citeauthoryear{{Barke}, {Wang}, {Esteban Delgado}, {Tr{\"o}bs}, {Heinzel}  \& {Danzmann}}{{Barke} et~al.}{2015}]{Barke2015}
{Barke} S.,  {Wang} Y.,  {Esteban Delgado} J.~J.,  {Tr{\"o}bs} M.,  {Heinzel} G.,   {Danzmann} K.,  2015, \mn@doi [Classical and Quantum Gravity] {10.1088/0264-9381/32/9/095004}, 32, 095004

\bibitem[\protect\citeauthoryear{Bayle \& Hartwig}{Bayle \& Hartwig}{2023}]{PhysRevD.107.083019}
Bayle J.-B.,  Hartwig O.,  2023, \mn@doi [Phys. Rev. D] {10.1103/PhysRevD.107.083019}, 107, 083019

\bibitem[\protect\citeauthoryear{Bellamy, Cairns  \& Smith}{Bellamy et~al.}{2005}]{voyagerturbulence}
Bellamy B.~R.,  Cairns I.~H.,   Smith C.~W.,  2005, \mn@doi [Journal of Geophysical Research: Space Physics] {https://doi.org/10.1029/2004JA010952}, 110

\bibitem[\protect\citeauthoryear{{Blas} \& {Jenkins}}{{Blas} \& {Jenkins}}{2022}]{2022PhRvL.128j1103B}
{Blas} D.,  {Jenkins} A.~C.,  2022, \mn@doi [\prl] {10.1103/PhysRevLett.128.101103}, \href {https://ui.adsabs.harvard.edu/abs/2022PhRvL.128j1103B} {128, 101103}

\bibitem[\protect\citeauthoryear{{Boone} \& {McQuinn}}{{Boone} \& {McQuinn}}{2023}]{2023ApJ...947L..23B}
{Boone} K.,  {McQuinn} M.,  2023, \mn@doi [\apjl] {10.3847/2041-8213/acc947}, \href {https://ui.adsabs.harvard.edu/abs/2023ApJ...947L..23B} {947, L23}

\bibitem[\protect\citeauthoryear{{Burt} et~al.,}{{Burt} et~al.}{2021}]{Burt2020}
{Burt} E.~A.,  et~al., 2021, \mn@doi [\nat] {10.1038/s41586-021-03571-7}, \href {https://ui.adsabs.harvard.edu/abs/2021Natur.595...43B} {595, 43}

\bibitem[\protect\citeauthoryear{{Colmenares}, {Battat}, {Gonzales}, {Murphy}  \& {Sabhlok}}{{Colmenares} et~al.}{2023}]{2023PASP..135j4503C}
{Colmenares} N.~R.,  {Battat} J.~B.~R.,  {Gonzales} D.~P.,  {Murphy} T.~W.,   {Sabhlok} S.,  2023, \mn@doi [\pasp] {10.1088/1538-3873/acf787}, \href {https://ui.adsabs.harvard.edu/abs/2023PASP..135j4503C} {135, 104503}

\bibitem[\protect\citeauthoryear{{Colpi} et~al.,}{{Colpi} et~al.}{2024}]{LISA}
{Colpi} M.,  et~al., 2024, \mn@doi [arXiv e-prints] {10.48550/arXiv.2402.07571}, \href {https://ui.adsabs.harvard.edu/abs/2024arXiv240207571C} {p. arXiv:2402.07571}

\bibitem[\protect\citeauthoryear{{Cornish} \& {Robson}}{{Cornish} \& {Robson}}{2017}]{2017JPhCS.840a2024C}
{Cornish} N.,  {Robson} T.,  2017, in Journal of Physics Conference Series. IOP, p. 012024 (\mn@eprint {arXiv} {1703.09858}), \mn@doi{10.1088/1742-6596/840/1/012024}

\bibitem[\protect\citeauthoryear{Crosta, Lattanzi, Le~Poncin-Lafitte, Gai, Zhaoxiang  \& Vecchiato}{Crosta et~al.}{2024}]{astrometryGW}
Crosta M.,  Lattanzi M.~G.,  Le~Poncin-Lafitte C.,  Gai M.,  Zhaoxiang Q.,   Vecchiato A.,  2024, \mn@doi [Scientific Reports] {10.1038/s41598-024-55671-9}, 14, 5074

\bibitem[\protect\citeauthoryear{{EPTA Collaboration} et~al.,}{{EPTA Collaboration} et~al.}{2023}]{2023A&A...678A..50E}
{EPTA Collaboration} et~al., 2023, \mn@doi [\aap] {10.1051/0004-6361/202346844}, \href {https://ui.adsabs.harvard.edu/abs/2023A&A...678A..50E} {678, A50}

\bibitem[\protect\citeauthoryear{{Estabrook} \& {Wahlquist}}{{Estabrook} \& {Wahlquist}}{1975}]{1975GReGr...6..439E}
{Estabrook} F.~B.,  {Wahlquist} H.~D.,  1975, \mn@doi [General Relativity and Gravitation] {10.1007/BF00762449}, \href {https://ui.adsabs.harvard.edu/abs/1975GReGr...6..439E} {6, 439}

\bibitem[\protect\citeauthoryear{{Faller}, {Bender}, {Hall}, {Hils}  \& {Vincent}}{{Faller} et~al.}{1985}]{1985ESASP.226..157F}
{Faller} J.~E.,  {Bender} P.~L.,  {Hall} J.~L.,  {Hils} D.,   {Vincent} M.~A.,  1985, in {Longdon} N.,  {Melita} O.,  eds,  ESA Special Publication Vol. 226, Kilometric Optical Arrays in Space. pp 157--163

\bibitem[\protect\citeauthoryear{{Fedderke}, {Graham}  \& {Rajendran}}{{Fedderke} et~al.}{2021}]{2021PhRvD.103j3017F}
{Fedderke} M.~A.,  {Graham} P.~W.,   {Rajendran} S.,  2021, \mn@doi [\prd] {10.1103/PhysRevD.103.103017}, \href {https://ui.adsabs.harvard.edu/abs/2021PhRvD.103j3017F} {103, 103017}

\bibitem[\protect\citeauthoryear{{Fedderke}, {Graham}  \& {Rajendran}}{{Fedderke} et~al.}{2022a}]{2022PhRvD.105j3018F}
{Fedderke} M.~A.,  {Graham} P.~W.,   {Rajendran} S.,  2022a, \mn@doi [\prd] {10.1103/PhysRevD.105.103018}, \href {https://ui.adsabs.harvard.edu/abs/2022PhRvD.105j3018F} {105, 103018}

\bibitem[\protect\citeauthoryear{{Fedderke}, {Graham}, {Macintosh}  \& {Rajendran}}{{Fedderke} et~al.}{2022b}]{2022PhRvD.106b3002F}
{Fedderke} M.~A.,  {Graham} P.~W.,  {Macintosh} B.,   {Rajendran} S.,  2022b, \mn@doi [\prd] {10.1103/PhysRevD.106.023002}, \href {https://ui.adsabs.harvard.edu/abs/2022PhRvD.106b3002F} {106, 023002}

\bibitem[\protect\citeauthoryear{{Folkner}}{{Folkner}}{2011}]{Folkner2011}
{Folkner} W.~M.,  2011, Technical report, A non-drag-free gravitational wave mission architecture, \url {https://pcos.gsfc.nasa.gov/studies/architecting/grav-wave/submissions/GWRFI-0003-Folkner.pdf}.
\url {https://pcos.gsfc.nasa.gov/studies/architecting/grav-wave/submissions/GWRFI-0003-Folkner.pdf}

\bibitem[\protect\citeauthoryear{{Fountain} et~al.,}{{Fountain} et~al.}{2008}]{2008SSRv..140...23F}
{Fountain} G.~H.,  et~al., 2008, \mn@doi [\ssr] {10.1007/s11214-008-9374-8}, \href {https://ui.adsabs.harvard.edu/abs/2008SSRv..140...23F} {140, 23}

\bibitem[\protect\citeauthoryear{{Fr{\"o}hlich} \& {Lean}}{{Fr{\"o}hlich} \& {Lean}}{2004}]{2004A&ARv..12..273F}
{Fr{\"o}hlich} C.,  {Lean} J.,  2004, \mn@doi [\aapr] {10.1007/s00159-004-0024-1}, \href {https://ui.adsabs.harvard.edu/abs/2004A&ARv..12..273F} {12, 273}

\bibitem[\protect\citeauthoryear{{Gr{\"u}n}, {Gustafson}, {Mann}, {Baguhl}, {Morfill}, {Staubach}, {Taylor}  \& {Zook}}{{Gr{\"u}n} et~al.}{1994}]{1994A&A...286..915G}
{Gr{\"u}n} E.,  {Gustafson} B.,  {Mann} I.,  {Baguhl} M.,  {Morfill} G.~E.,  {Staubach} P.,  {Taylor} A.,   {Zook} H.~A.,  1994, \aap, \href {https://ui.adsabs.harvard.edu/abs/1994A&A...286..915G} {286, 915}

\bibitem[\protect\citeauthoryear{{Gr{\"u}n}, {Kr{\"u}ger}  \& {Landgraf}}{{Gr{\"u}n} et~al.}{2000}]{2000mbos.work...99G}
{Gr{\"u}n} E.,  {Kr{\"u}ger} H.,   {Landgraf} M.,  2000, in {Fitzsimmons} A.,  {Jewitt} D.,   {West} R.~M.,  eds, Minor Bodies in the Outer Solar System. p.~99 (\mn@eprint {arXiv} {astro-ph/9902036}), \mn@doi{10.1007/10651968_13}

\bibitem[\protect\citeauthoryear{Harris et~al.,}{Harris et~al.}{2020}]{numpy}
Harris C.~R.,  et~al., 2020, \mn@doi [Nature] {10.1038/s41586-020-2649-2}, 585, 357

\bibitem[\protect\citeauthoryear{Heinzel, \'Alvarez-Vizoso, Dovale-\'Alvarez  \& Wiesner}{Heinzel et~al.}{2024}]{PhysRevD.110.042002}
Heinzel G.,  \'Alvarez-Vizoso J.,  Dovale-\'Alvarez M.,   Wiesner K.,  2024, \mn@doi [Phys. Rev. D] {10.1103/PhysRevD.110.042002}, 110, 042002

\bibitem[\protect\citeauthoryear{{High Frequency Electronics}}{{High Frequency Electronics}}{2023}]{rf_lna_technology_landscape}
{High Frequency Electronics} 2023, High Frequency Electronics

\bibitem[\protect\citeauthoryear{Hoang et~al.,}{Hoang et~al.}{2023}]{m2picclock}
Hoang T.~M.,  et~al., 2023, \mn@doi [Scientific Reports] {10.1038/s41598-023-36411-x}, 13, 10629

\bibitem[\protect\citeauthoryear{Hunter}{Hunter}{2007}]{Hunter:2007}
Hunter J.~D.,  2007, \mn@doi [Computing in Science \& Engineering] {10.1109/MCSE.2007.55}, 9, 90

\bibitem[\protect\citeauthoryear{{IEEE Standards Association}}{{IEEE Standards Association}}{2009}]{IEEEclock}
{IEEE Standards Association} 2009, \mn@doi [IEEE Std Std 1139-2008] {10.1109/IEEESTD.2008.4797525}, pp c1--35

\bibitem[\protect\citeauthoryear{{Kardashev}, {Kovalev}  \& {Kellermann}}{{Kardashev} et~al.}{2012}]{2012RSB...343...22K}
{Kardashev} N.~S.,  {Kovalev} Y.~Y.,   {Kellermann} K.~I.,  2012, \mn@doi [The Radio Science Bulletin No 343] {10.48550/arXiv.1303.5200}, \href {https://ui.adsabs.harvard.edu/abs/2012RSB...343...22K} {343, 22}

\bibitem[\protect\citeauthoryear{Kluyver et~al.,}{Kluyver et~al.}{2016}]{kluyver2016jupyter}
Kluyver T.,  et~al., 2016, in ELPUB. pp 87--90

\bibitem[\protect\citeauthoryear{Kolkowitz, Pikovski, Langellier, Lukin, Walsworth  \& Ye}{Kolkowitz et~al.}{2016}]{PhysRevD.94.124043}
Kolkowitz S.,  Pikovski I.,  Langellier N.,  Lukin M.~D.,  Walsworth R.~L.,   Ye J.,  2016, \mn@doi [Phys. Rev. D] {10.1103/PhysRevD.94.124043}, 94, 124043

\bibitem[\protect\citeauthoryear{LAI}{LAI}{2012}]{ba3eaa77-fc94-3442-be38-0f8c614f5d3e}
LAI S.~T.,  2012, Fundamentals of Spacecraft Charging: Spacecraft Interactions with Space Plasmas.
Princeton University Press, \url {http://www.jstor.org/stable/j.ctvcm4j2n}

\bibitem[\protect\citeauthoryear{{Larson}, {Hiscock}  \& {Hellings}}{{Larson} et~al.}{2000}]{larson01}
{Larson} S.~L.,  {Hiscock} W.~A.,   {Hellings} R.~W.,  2000, \mn@doi [Phys. Rev. D] {10.1103/PhysRevD.62.062001}, 62, 062001

\bibitem[\protect\citeauthoryear{Larson, Hellings  \& Hiscock}{Larson et~al.}{2002}]{PhysRevD.66.062001}
Larson S.~L.,  Hellings R.~W.,   Hiscock W.~A.,  2002, \mn@doi [Phys. Rev. D] {10.1103/PhysRevD.66.062001}, 66, 062001

\bibitem[\protect\citeauthoryear{{Loeb} \& {Maoz}}{{Loeb} \& {Maoz}}{2015}]{2015arXiv150100996L}
{Loeb} A.,  {Maoz} D.,  2015, \mn@doi [arXiv e-prints] {10.48550/arXiv.1501.00996}, \href {https://ui.adsabs.harvard.edu/abs/2015arXiv150100996L} {p. arXiv:1501.00996}

\bibitem[\protect\citeauthoryear{{Lu}, {Wang}  \& {Xiao}}{{Lu} et~al.}{2024}]{2024arXiv240712920L}
{Lu} Z.,  {Wang} L.-T.,   {Xiao} H.,  2024, \mn@doi [arXiv e-prints] {10.48550/arXiv.2407.12920}, \href {https://ui.adsabs.harvard.edu/abs/2024arXiv240712920L} {p. arXiv:2407.12920}

\bibitem[\protect\citeauthoryear{Marini \& C.~W.~Murray}{Marini \& C.~W.~Murray}{1973}]{Marini_1973}
Marini J.~W.,  C.~W.~Murray J.,  1973, Technical Report NASA-TM-X-70555, Correction of Laser Range Tracking Data for Atmospheric Refraction at Elevations Above 10 Degrees, \url {https://ntrs.nasa.gov/citations/19740007037}.
NASA Goddard Space Flight Center, \url {https://ntrs.nasa.gov/citations/19740007037}

\bibitem[\protect\citeauthoryear{{Marsat}, {Baker}  \& {Canton}}{{Marsat} et~al.}{2021}]{2021PhRvD.103h3011M}
{Marsat} S.,  {Baker} J.~G.,   {Canton} T.~D.,  2021, \mn@doi [\prd] {10.1103/PhysRevD.103.083011}, \href {https://ui.adsabs.harvard.edu/abs/2021PhRvD.103h3011M} {103, 083011}

\bibitem[\protect\citeauthoryear{{Martens}, {Khan}  \& {Bayle}}{{Martens} et~al.}{2023}]{2023CQGra..40s5022M}
{Martens} W.,  {Khan} M.,   {Bayle} J.~B.,  2023, \mn@doi [Classical and Quantum Gravity] {10.1088/1361-6382/acf3c7}, \href {https://ui.adsabs.harvard.edu/abs/2023CQGra..40s5022M} {40, 195022}

\bibitem[\protect\citeauthoryear{{McKenzie} et~al.,}{{McKenzie} et~al.}{2011}]{McKenzie2011}
{McKenzie} K.,  et~al., 2011, Technical report, LAGRANGE: A Space-Based Gravitational-Wave Detector with Geometric Suppression of Spacecraft Noise.
Jet Propulsion Laboratory, California Institute of Technology, Pasadena, California

\bibitem[\protect\citeauthoryear{{Misra} \& {Enge}}{{Misra} \& {Enge}}{2012}]{misraenge}
{Misra} P.,  {Enge} P.,  2012, {Global positioning system: signals, measurements, and performance. Rev. 2nd ed. Lincoln, Mass.: Ganga-Jamuna Press.}

\bibitem[\protect\citeauthoryear{{Mueller} et~al.,}{{Mueller} et~al.}{2019}]{2019BAAS...51g.243M}
{Mueller} G.,  et~al., 2019, in Bulletin of the American Astronomical Society. p.~243 (\mn@eprint {arXiv} {1907.11305}), \mn@doi{10.48550/arXiv.1907.11305}

\bibitem[\protect\citeauthoryear{{NASA}}{{NASA}}{2020}]{nasa2020sls}
{NASA} 2020, Technical Report 20205007434, {The Space Launch System for Science: A Launch Capability for the Next Decadal Survey and Beyond}, \url {https://ntrs.nasa.gov/api/citations/20205007434/downloads/SLS-PlanetaryDecadal-FINAL-LRP1%20091420.pdf}.
{NASA}, \url {https://ntrs.nasa.gov/api/citations/20205007434/downloads/SLS-PlanetaryDecadal-FINAL-LRP1%20091420.pdf}

\bibitem[\protect\citeauthoryear{{Narayan}}{{Narayan}}{1992}]{1992RSPTA.341..151N}
{Narayan} R.,  1992, \mn@doi [Philosophical Transactions of the Royal Society of London Series A] {10.1098/rsta.1992.0090}, \href {https://ui.adsabs.harvard.edu/abs/1992RSPTA.341..151N} {341, 151}

\bibitem[\protect\citeauthoryear{Neronov, Roper~Pol, Caprini  \& Semikoz}{Neronov et~al.}{2021}]{PhysRevD.103.L041302}
Neronov A.,  Roper~Pol A.,  Caprini C.,   Semikoz D.,  2021, \mn@doi [Phys. Rev. D] {10.1103/PhysRevD.103.L041302}, 103, L041302

\bibitem[\protect\citeauthoryear{{Ni}}{{Ni}}{2010}]{2010MPLA...25..922N}
{Ni} W.-T.,  2010, \mn@doi [Modern Physics Letters A] {10.1142/S0217732310000071}, \href {https://ui.adsabs.harvard.edu/abs/2010MPLA...25..922N} {25, 922}

\bibitem[\protect\citeauthoryear{{Nissanke}, {Vallisneri}, {Nelemans}  \& {Prince}}{{Nissanke} et~al.}{2012}]{2012ApJ...758..131N}
{Nissanke} S.,  {Vallisneri} M.,  {Nelemans} G.,   {Prince} T.~A.,  2012, \mn@doi [\apj] {10.1088/0004-637X/758/2/131}, \href {https://ui.adsabs.harvard.edu/abs/2012ApJ...758..131N} {758, 131}

\bibitem[\protect\citeauthoryear{{Njoku} et~al.,}{{Njoku} et~al.}{2014}]{SMAP_Handbook}
{Njoku} E.,  et~al., 2014, SMAP Handbook.
Jet Propulsion Laboratory, California Institute of Technology

\bibitem[\protect\citeauthoryear{Numata, Kemery  \& Camp}{Numata et~al.}{2004}]{PhysRevLett.93.250602}
Numata K.,  Kemery A.,   Camp J.,  2004, \mn@doi [Phys. Rev. Lett.] {10.1103/PhysRevLett.93.250602}, 93, 250602

\bibitem[\protect\citeauthoryear{Oelker et~al.,}{Oelker et~al.}{2019}]{greatatomicclocks}
Oelker E.,  et~al., 2019, \mn@doi [Nature Photonics] {10.1038/s41566-019-0493-4}, 13, 714

\bibitem[\protect\citeauthoryear{{Perez} \& {Granger}}{{Perez} \& {Granger}}{2007}]{2007CSE.....9c..21P}
{Perez} F.,  {Granger} B.~E.,  2007, \mn@doi [Computing in Science and Engineering] {10.1109/MCSE.2007.53}, \href {https://ui.adsabs.harvard.edu/abs/2007CSE.....9c..21P} {9, 21}

\bibitem[\protect\citeauthoryear{{Reardon} et~al.,}{{Reardon} et~al.}{2023}]{2023ApJ...951L...6R}
{Reardon} D.~J.,  et~al., 2023, \mn@doi [\apjl] {10.3847/2041-8213/acdd02}, \href {https://ui.adsabs.harvard.edu/abs/2023ApJ...951L...6R} {951, L6}

\bibitem[\protect\citeauthoryear{{Robson}, {Cornish}  \& {Liu}}{{Robson} et~al.}{2019}]{2019CQGra..36j5011R}
{Robson} T.,  {Cornish} N.~J.,   {Liu} C.,  2019, \mn@doi [Classical and Quantum Gravity] {10.1088/1361-6382/ab1101}, \href {https://ui.adsabs.harvard.edu/abs/2019CQGra..36j5011R} {36, 105011}

\bibitem[\protect\citeauthoryear{Rosado}{Rosado}{2011}]{PhysRevD.84.084004}
Rosado P.~A.,  2011, \mn@doi [Phys. Rev. D] {10.1103/PhysRevD.84.084004}, 84, 084004

\bibitem[\protect\citeauthoryear{Sambridge et~al.,}{Sambridge et~al.}{2023}]{PhysRevLett.131.193804}
Sambridge C.~S.,  et~al., 2023, \mn@doi [Phys. Rev. Lett.] {10.1103/PhysRevLett.131.193804}, 131, 193804

\bibitem[\protect\citeauthoryear{Sambridge, Valliyakalayil  \& McKenzie}{Sambridge et~al.}{2024}]{rs16193598}
Sambridge C.~S.,  Valliyakalayil J.~T.,   McKenzie K.,  2024, \mn@doi [Remote Sensing] {10.3390/rs16193598}, 16

\bibitem[\protect\citeauthoryear{{Sesana} et~al.,}{{Sesana} et~al.}{2021}]{2021ExA....51.1333S}
{Sesana} A.,  et~al., 2021, \mn@doi [Experimental Astronomy] {10.1007/s10686-021-09709-9}, \href {https://ui.adsabs.harvard.edu/abs/2021ExA....51.1333S} {51, 1333}

\bibitem[\protect\citeauthoryear{Shaddock, Tinto, Estabrook  \& Armstrong}{Shaddock et~al.}{2003}]{TDI_with_motion}
Shaddock D.~A.,  Tinto M.,  Estabrook F.~B.,   Armstrong J.~W.,  2003, \mn@doi [Phys. Rev. D] {10.1103/PhysRevD.68.061303}, 68, 061303

\bibitem[\protect\citeauthoryear{Smith}{Smith}{1989}]{SMITH1989159}
Smith E.,  1989, \mn@doi [Advances in Space Research] {https://doi.org/10.1016/0273-1177(89)90110-5}, 9, 159

\bibitem[\protect\citeauthoryear{Stacey, Barwood, Spampinato, Tsoulos, Robinson, Gaynor  \& Gill}{Stacey et~al.}{2023}]{10.1117/12.2691441}
Stacey J.,  Barwood G.,  Spampinato A.,  Tsoulos P.,  Robinson C.,  Gaynor P.,   Gill P.,  2023, in Minoglou K.,  Karafolas N.,   Cugny B.,  eds,  Proceedings of SPIE Vol. 12777, International Conference on Space Optics - ICSO 2022. SPIE, p. 127777F, \mn@doi{10.1117/12.2691441}, \url {https://doi.org/10.1117/12.2691441}

\bibitem[\protect\citeauthoryear{Taylor}{Taylor}{2016}]{spacecommunications}
Taylor J.,  2016, Deep Space Communications.
John Wiley \& Sons, Incorporated, Newark, UNITED STATES, \url {http://ebookcentral.proquest.com/lib/washington/detail.action?docID=4648722}

\bibitem[\protect\citeauthoryear{Tinto \& Dhurandhar}{Tinto \& Dhurandhar}{2014}]{TDIreview}
Tinto M.,  Dhurandhar S.~V.,  2014, \mn@doi [Living Reviews in Relativity] {10.12942/lrr-2014-6}, 17, 6

\bibitem[\protect\citeauthoryear{Tomio et~al.,}{Tomio et~al.}{2024}]{Tomio_2024}
Tomio H.,  et~al., 2024, in Navarro R.,  Jedamzik R.,  eds, Advances in Optical and Mechanical Technologies for Telescopes and Instrumentation VI. SPIE, p.~24, \mn@doi{10.1117/12.3020732}, \url {http://dx.doi.org/10.1117/12.3020732}

\bibitem[\protect\citeauthoryear{{Vallisneri} et~al.,}{{Vallisneri} et~al.}{2020}]{2020ApJ...893..112V}
{Vallisneri} M.,  et~al., 2020, \mn@doi [\apj] {10.3847/1538-4357/ab7b67}, \href {https://ui.adsabs.harvard.edu/abs/2020ApJ...893..112V} {893, 112}

\bibitem[\protect\citeauthoryear{Van~Rossum \& Drake}{Van~Rossum \& Drake}{2009}]{python}
Van~Rossum G.,  Drake F.~L.,  2009, Python 3 Reference Manual.
CreateSpace, Scotts Valley, CA

\bibitem[\protect\citeauthoryear{{Wagg} \& {Broekgaarden}}{{Wagg} \& {Broekgaarden}}{2024}]{software-citation-station-paper}
{Wagg} T.,  {Broekgaarden} F.~S.,  2024, arXiv e-prints, \href {https://ui.adsabs.harvard.edu/abs/2024arXiv240604405W} {p. arXiv:2406.04405}

\bibitem[\protect\citeauthoryear{Wang, Calhoun, Kirk, Diener, Dick  \& Tjoelker}{Wang et~al.}{2005}]{Wang2005}
Wang R.~T.,  Calhoun M.~D.,  Kirk A.,  Diener W.~A.,  Dick G.~J.,   Tjoelker R.~L.,  2005, in 2005 Joint IEEE International Frequency Control Symposium (FCS) and Precise Time and Time Interval (PTTI) Systems and Applications Meeting. Vancouver, British Columbia, Canada, \url {https://ntrs.nasa.gov/search.jsp?R=20060044174}

\bibitem[\protect\citeauthoryear{Wang, Davydov  \& Rud}{Wang et~al.}{2021}]{Wang_2021}
Wang D.,  Davydov V.~V.,   Rud V.~Y.,  2021, \mn@doi [Journal of Physics: Conference Series] {10.1088/1742-6596/2086/1/012073}, 2086, 012073

\bibitem[\protect\citeauthoryear{{Wang}, {Pardo}, {Chang}  \& {Dor{\'e}}}{{Wang} et~al.}{2022}]{2022PhRvD.106h4006W}
{Wang} Y.,  {Pardo} K.,  {Chang} T.-C.,   {Dor{\'e}} O.,  2022, \mn@doi [\prd] {10.1103/PhysRevD.106.084006}, \href {https://ui.adsabs.harvard.edu/abs/2022PhRvD.106h4006W} {106, 084006}

\bibitem[\protect\citeauthoryear{Wissel, Wittchen, Schwarze, Hewitson, Heinzel  \& Halloin}{Wissel et~al.}{2022}]{PhysRevApplied.17.024025}
Wissel L.,  Wittchen A.,  Schwarze T.~S.,  Hewitson M.,  Heinzel G.,   Halloin H.,  2022, \mn@doi [Phys. Rev. Appl.] {10.1103/PhysRevApplied.17.024025}, 17, 024025

\bibitem[\protect\citeauthoryear{Wissel, Hartwig, Bayle, Staab, Fitzsimons, Hewitson  \& Heinzel}{Wissel et~al.}{2023}]{PhysRevApplied.20.014016}
Wissel L.,  Hartwig O.,  Bayle J.,  Staab M.,  Fitzsimons E.,  Hewitson M.,   Heinzel G.,  2023, \mn@doi [Phys. Rev. Appl.] {10.1103/PhysRevApplied.20.014016}, 20, 014016

\bibitem[\protect\citeauthoryear{Woo \& Guo}{Woo \& Guo}{2004}]{Woo_2004}
Woo R.,  Guo W.~H.,  2004, Technical Report~158A, A New Technique for Determining the Strength of Interstellar Scintillation, \url {https://tmo.jpl.nasa.gov/progress_report/42-158/158A.pdf}.
NASA Jet Propulsion Laboratory, \url {https://tmo.jpl.nasa.gov/progress_report/42-158/158A.pdf}

\bibitem[\protect\citeauthoryear{{Xiao}, {Dai}  \& {McQuinn}}{{Xiao} et~al.}{2024}]{2024PhRvD.110b3516X}
{Xiao} H.,  {Dai} L.,   {McQuinn} M.,  2024, \mn@doi [\prd] {10.1103/PhysRevD.110.023516}, \href {https://ui.adsabs.harvard.edu/abs/2024PhRvD.110b3516X} {110, 023516}

\bibitem[\protect\citeauthoryear{{Zwick} et~al.,}{{Zwick} et~al.}{2024}]{2024arXiv240602306Z}
{Zwick} L.,  et~al., 2024, \mn@doi [arXiv e-prints] {10.48550/arXiv.2406.02306}, \href {https://ui.adsabs.harvard.edu/abs/2024arXiv240602306Z} {p. arXiv:2406.02306}

\makeatother
\end{thebibliography}

\appendix

\section{Dispersion}
\label{sec:dispersion}

\begin{figure}
\begin{center}
\includegraphics[width=0.6\textwidth]{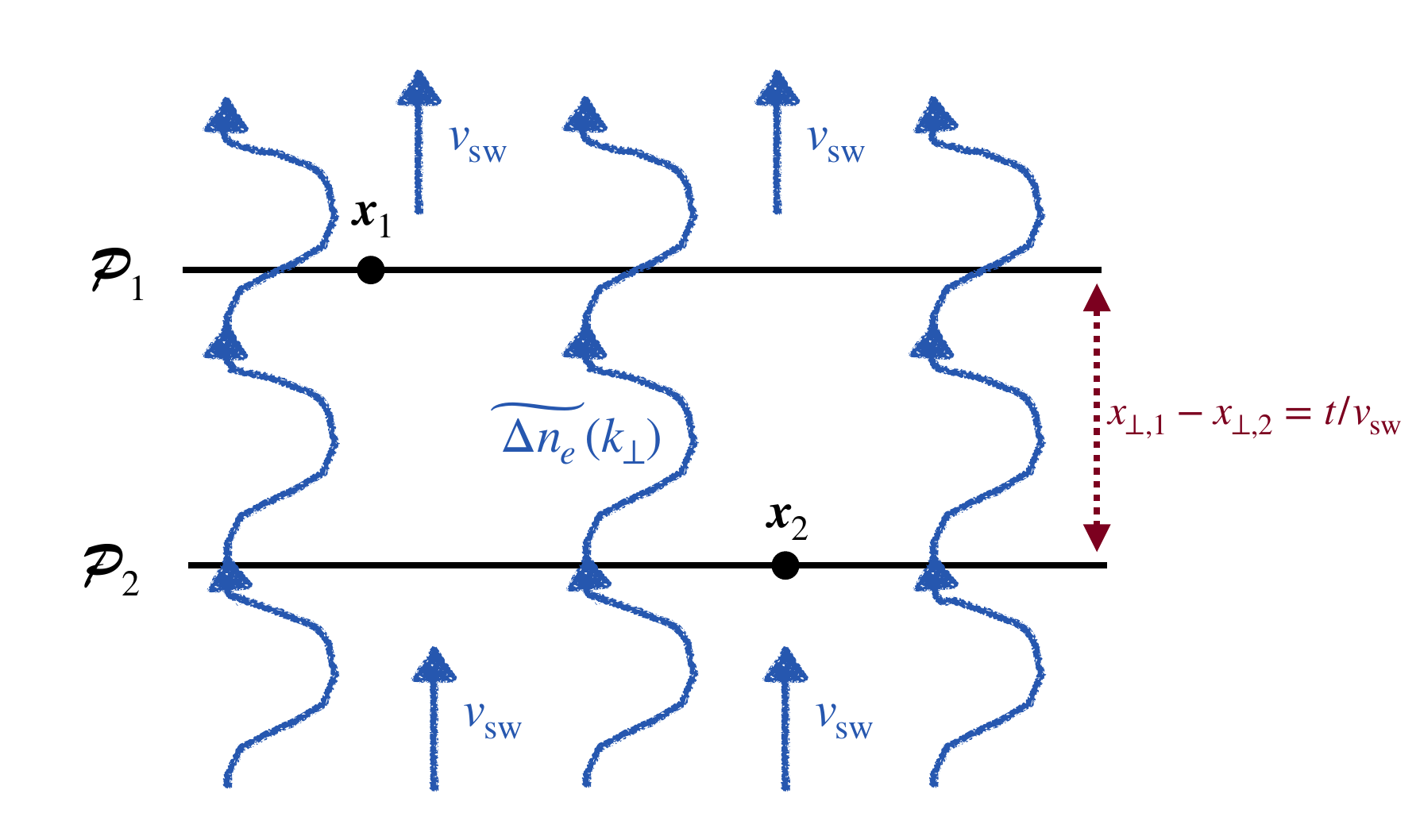}
\end{center}
\vspace{-1.cm}
\caption{Illustration of the setup of the calculation presented in Appendixes~\ref{sec:dispersion} and \ref{sec:refraction} for the dispersive delays owing to the interplanetary plasma.  Parallel paths ${\cal P}_1$ and ${\cal P}_2$ represent the arms linking two spacecraft at different times.  We calculate the phase-delay correlation between these paths separated by $x_{\perp 1}-x_{\perp 2}$ and oriented perpendicular to the direction of the solar wind.  We treat the solar wind as acting to translate inhomogeneities in the electron density with velocity $v_{\rm sw}$, mapping spatial correlations between paths ${\cal P}_1$ and ${\cal P}_2$ to the desired temporal correlations.  The spectra of electron density fluctuations are represented by $\widetilde{\Delta n_e}(\bfk)$, and only modes that have perpendicular orientations to the path contribute substantially to the correlations.  One such nearly perpendicular mode of electron density fluctuations is illustrated.\label{fig:dispersionvis}}
\end{figure}

An issue with radio observations that is not present for our laser setup is that dispersion in the interplanetary plasma will contribute phase noise.   Taking the density of the solar wind to be $n_e= 0.05 (r/10 {\rm \, AU})^{-2}\,$cm$^{-3}$ (e.g., \citealt{voyagerturbulence}), plasma dispersion leads to an error on the strain of
\begin{equation}
\delta h \sim c \kappa n_e  = 2.6\times 10^{-15} \left( \frac{r}{30~{\rm AU} }\right)^{-2} \left( \frac{\lambda}{1~\rm cm} \right)^{2},
\end{equation}
where $\kappa = \lambda^2 e^2/(2\pi m_e c^3)$ converts an electron column density to a phase delay.  Of course, the uniform solar wind signal does not look like gravitational waves.  We can calculate the inhomogeneous part in terms of the power spectrum of the solar wind electrons, $P_e$, by first calculating the temporal correlation function of the plasma delay between the two dishes.  Additionally, because the solar wind acts to radially translate the inhomogeneities, the correlation function of delays between parallel paths can be related to the temporal correlation function that we desire.   We approximate the paths along the arms as being at a fixed solar distance $r$ and perpendicular to the direction of the Sun.  We additionally assume that the perturbations along the path that the light travels do not change over the round-trip light-travel time for an arm, which means that our calculation will somewhat overestimate the effect for frequencies that satisfy $f \gtrsim c/(2 L)$.  Figure~\ref{fig:dispersionvis} illustrates the setup.

The correlation of the phase delays between two parallel paths is
\begin{align}
    \expectation{\Delta\tau_{d,1} \Delta\tau_{d,2}} 
     &= \kappa^2 \expectation{\int_{{\cal P}_1} dx_1 \Delta n_e(\bfx_1) \int_{{\cal P}_2} dx_2 \Delta n_e(\bfx_2)},
\end{align}
where $\Delta n_e$ is the 3D field of electron density fluctuations. Writing $\Delta n_e$ in terms of its Fourier transform, the expectation value becomes
\begin{align}
    \expectation{\Delta\tau_{d,1} \Delta\tau_{d,2}} = & \kappa^2 \int_{{\cal P}_1} dx_1 \int_{{\cal P}_2} dx_2 \iint \frac{d^3\bfk_1 d^3\bfk_2}{(2\pi)^6}  \expectation{\widetilde{\Delta n_e}(\bfk_1) \widetilde{\Delta n_e}(\bfk_2)^*} e^{i (\bfk_1 \cdot \bfx_1 - \bfk_2 \cdot \bfx_2)}, \\
    = & \kappa^2 \int_{{\cal P}_1} dx_1 \int_{{\cal P}_2} dx_2 \int \frac{d^3\bfk}{(2\pi)^3} P_e(k) e^{i \bfk \cdot (\bfx_1 - \bfx_2)},\label{eqn:tautdPe}
\end{align}
where we have defined the electron density power spectrum as
    $\expectation{\widetilde{\Delta n_e}(\bfk_1) \widetilde{\Delta n_e}(\bfk_2)^*} = (2 \pi)^3 P_e(k_1) \delta^D(\bfk_1-\bfk_2)$.
In the `Limber approximation limit' that applies when the integral's support comes from $k \gg 2\pi/L$, the integral over the line-of-sight wavevector along ${\cal P}_1$ can be approximated as a $\delta$-function.  The $\delta$-function can then be used to eliminate the integral over the wavevector along the path.  The remaining integral along ${\cal P}_2$ evaluates to its length, $L$, such that equation~(\ref{eqn:tautdPe}) reduces to
\begin{align}
    \expectation{\Delta\tau_{d,1} \Delta\tau_{d,2}} \approx \kappa^2 L \int \frac{d^2\bfk_\perp}{(2\pi)^2} P_e(k) e^{i \bfk_\perp \cdot (\bfx_{\perp, 1} - \bfx_{\perp, 2})}, 
        \label{eqn:vartauintegral}
\end{align}
where $\bfx_{\perp, 1} - \bfx_{\perp, 2}$ is the minimum separation between the two parallel paths.

The half-bandwidth temporal power of phase delay fluctuations is given by twice the temporal Fourier transform of $\expectation{\Delta\tau_{d,1} \Delta\tau_{d,2}}$, using that time is related to position by
$\bfx_{\perp,1} - \bfx_{\perp,2} = t/v_{\rm sw} \hatn$, where $\hatn$ is the direction of the solar wind as well as the direction perpendicular to our two parallel paths.  Thus,
\begin{eqnarray}
P_{\Delta \tau_d}(f) &=& 2 \int dt \, e^{-i \omega t}  \expectation{\Delta\tau_{d,1} \Delta\tau_{d,2}} = 2 \kappa^2 L \int \frac{d^2\bfk_\perp}{(2\pi)^2} P_e(k) \; (2\pi) \delta^D\left(\omega - v_{\rm sw}\bfk_\perp\cdot \hatn \right), \nonumber \\
&=& \frac{2 \kappa^2 L}{v_{\rm sw}} \int_{-\infty}^{\infty} \frac{d k_x}{(2\pi)} P_e\left(\sqrt{k_x^2+  \frac{\omega^2}{v_{\rm sw}^2}} \right), \label{eqn:Ptau2}\\
& \approx &1.5 \, \kappa^2 \, L \frac{v_{\rm sw}}{\omega} P_e^{\rm 1D}(f, r),\label{eqn:Ptaufinal}
\end{eqnarray}
 where $f \equiv \omega/2\pi$ and $k_x$ is the component of the wavevector perpendicular to the solar wind.  Equation~(\ref{eqn:Ptaufinal}) used that the 1D density power that is measured by, e.g., Voyager 2 is related to the 3D in the above expression by
\begin{equation}
P_e^{\rm 1D}(f) = \frac{2}{v_{\rm sw}}\int_{0}^{\infty} \frac{d k}{(2\pi)}  k P_e\left(\sqrt{k^2+  \frac{\omega^2}{v_{\rm sw}^2}}  \right),\label{eqn:Pe1D}
\end{equation}
where the factor of $2$ is because this is the half-bandwidth power.  Since the integral that yields $P_e^{\rm 1D}(f)$ is somewhat different than that in equation~(\ref{eqn:Ptau2}), to determine the numerical coefficient in equation~(\ref{eqn:Ptaufinal}) we assumed the scaling $P_e^{\rm 1D} \propto f^{-1.5}$ as motivated in \S~\ref{ss:solarwind}.  However, we find this coefficient depends weakly on what power-law index is assumed over relevant indices.\footnote{The coefficient changes from $1.5$ to $1.6$ if we use $P_e^{\rm 1D} \propto f^{-2}$, which may be more appropriate at $f \gtrsim 10^{-5}$Hz.}

 


We can convert this to fluctuations in phase as
\begin{equation}
\widetilde{\phi}_{\rm disp, rms} =  \frac{c}{2\pi \lambda} \sqrt{4 P_{\Delta \tau_d}(\omega)} ,
\label{eqn:phidisp}
\end{equation}
where the four is because the light travels out and back for each arm, probing essentially the same electron field for $f \lesssim c/(2L)$.  At higher frequencies, the return path will see different electrons such that we should replace $4\rightarrow 2$ -- a correction that we ignore. 

Converting the phase error to the error on the gravitational wave strain yields
\begin{eqnarray}
\widetilde{h}_{\rm disp, rms} 
                     &=& 4 \times 10^{-15} \, \text{Hz}^{-1/2} ~R(f)^{-1/2} \left( \frac{L}{30 \, {\rm AU}}\right)^{-1/2} \left( \frac{r}{30\,{\rm AU} }\right)^{-1.5} \nonumber \\
                     &&\times\left( \frac{f}{10^{-5} {\rm Hz}} \right)^{-(\beta+1)/2} \left( \frac{P_{e,0}^{\rm 1D}}{{10^{11.1} {\rm m}^{-6} {\rm Hz}^{-1}}} \right)^{1/2} \left( \frac{\lambda}{ 1~{\rm cm}} \right)^2, \label{eqn:hdisphrms}
\end{eqnarray}
where we are using the form of the 1D solar wind density power given by equation~(\ref{eqn:swdensitypower}), noting that $\beta \approx 2$ for $f <10^{-5}~$Hz  (\S~\ref{ss:solarwind}).  We include the transfer function $R(f)$ to convert from the long wavelength limit (cf. eqn.~\ref{eqn:Sh}). 

\begin{figure}
\begin{center}
\includegraphics[width=1.0\textwidth]{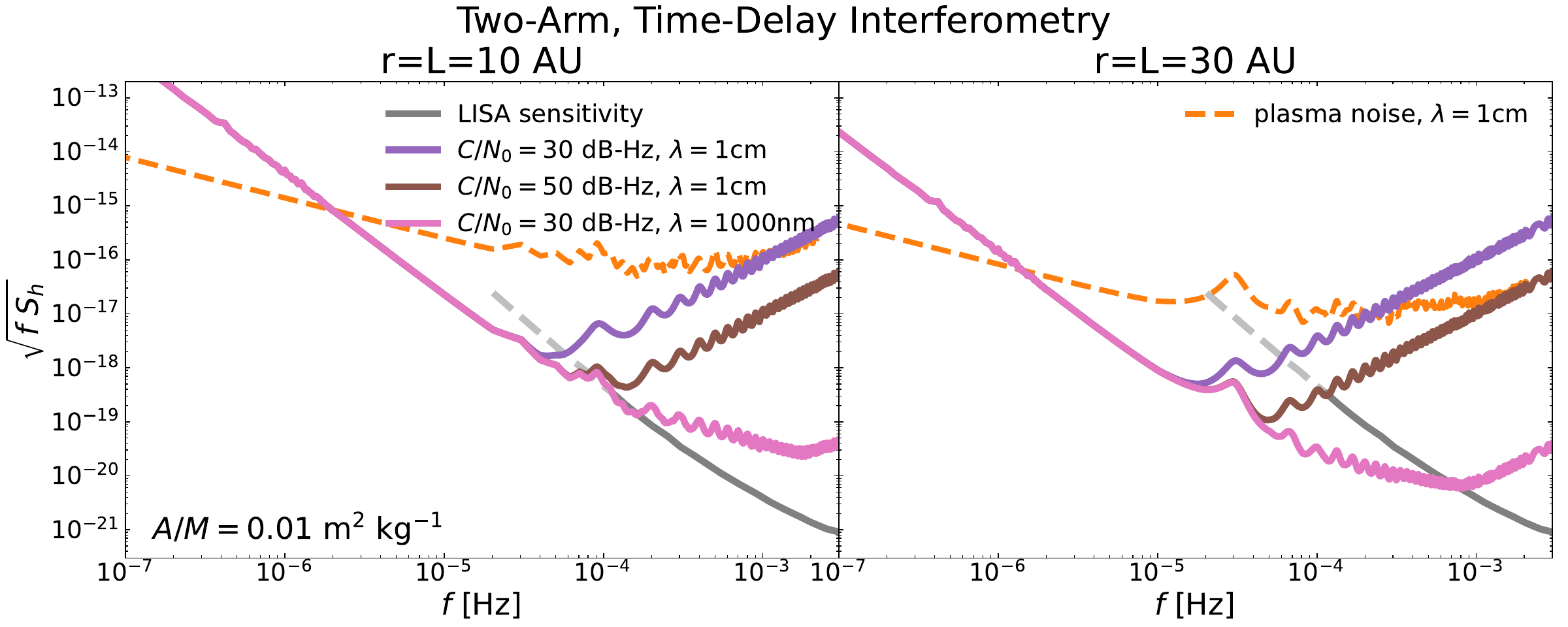}
\end{center}
\caption{Effect of dispersion for an interferometer operating at $\lambda =1\,$cm. The dashed curve shows the strain noise power we estimate from plasma dispersion.  The other curves show the total noise once correcting for plasma dispersion with ${\cal A} =1.5$ in the time-delay interferometer setup (the same curves as in Fig.~\ref{fig:totalsensitivity}). The left panel shows the case $r= L=10\;$AU, and the right panel $r= L=30\;$AU with the approximations discussed in \S~\ref{sec:dispersion}. The estimates in this figure use equation~(\ref{eqn:phidisp}) and the electron density power spectrum of the solar wind measured by Voyager 2 \citep{voyagerturbulence}.\label{fig:dedispersion}}
\end{figure}

Figure~\ref{fig:dedispersion} shows the effect of dispersion for an interferometer operating at $\lambda =1\,$cm.  The dashed curves show the strain noise power we estimate from plasma dispersion.  The other curves are our predictions for the noise in the time-delay interferometry configuration once correcting for plasma dispersion by transmitting at two frequencies (see below; these are the same curves as in Fig.~\ref{fig:totalsensitivity}). The left panel shows $r= L=10 \,$AU, and the right panel $r= L=30\,$AU assuming the spacecraft separation is nearly orthogonal to the radial direction so that the above calculations apply.  
  Rather than assuming a power-law scaling as in equation~(\ref{eqn:hdisphrms}), the estimates in this figure use equations~(\ref{eqn:Ptaufinal}) and (\ref{eqn:phidisp}) as well as the 1D electron density power spectrum of the solar wind measured in \citet{voyagerturbulence} using Voyager 2 data.\footnote{Since our derivation of equation~(\ref{eqn:Ptaufinal}) assumed a power law, it is not rigorous to use the measured 1D electron density power spectrum, $P^{\rm 1D}_e(f)$:  The different wavenumber weighting of the integrand over $P_e$ that yields the measured 1D electron density spectrum (eqn.~\ref{eqn:Pe1D}) and the phase noise delay power spectrum (eqn.~\ref{eqn:Ptaufinal}) means that the features in the measured $P^{\rm 1D}_e(f)$ should be somewhat distorted in the phase noise.}   This figure shows that the effect of dispersion is large and, if uncorrected, would limit the sensitivity of our concepts at intermediate frequencies.

While the phase noise due to the plasma can limit the sensitivity, it can be essentially eliminated by broadcasting at two wavelengths, $\lambda_1$ and $\lambda_2$, and then applying the following estimator for the inter-spacecraft displacement:  
\begin{equation}
\widehat{\Delta x} = \frac{\lambda_1 \lambda_2}{2\pi (\lambda_1^2 - \lambda_2^2)}\left(\lambda_1\phi_{2} -\lambda_2\phi_{1}  \right).
\label{eqn:dedisp}
\end{equation}
Accelerations and gravitational waves lead to the same $\Delta x$ as the previous single-phase estimator.  Therefore, this estimator's gravitational wave strain sensitivity is not affected when accelerations set the noise.  However, this estimator does come with the cost of an increased error when radiometer noise is important.  When radiometer noise dominates, if we assume $\delta \phi_j \propto \lambda_j$ as applies if  $A_{\rm eff}$ and $T_{\rm sys}$ are the same for the links at both $\lambda_1$ and $\lambda_2$, then phase errors are mapped to total errors as $\sigma_{\Delta x} = \sqrt{2} \lambda_1 \lambda_2^2/(2\pi |\lambda_1^2 - \lambda_2^2|) \, \sigma_{\phi_1}$, using that $\sigma_{\phi_2} = \lambda_2/\lambda_1 \sigma_{\phi_1}$,  rather than $\sigma_{\Delta x} =  \lambda_1/(2\pi )\, \sigma_{ \phi_1}$ as for the single-phase estimator.  Thus, it results in an increase in the estimator error by the factor $\sqrt{2} \lambda_2^2/|\lambda_2^2-\lambda_1^2|$.  If $\lambda_1 = 1\,$cm and $\lambda_2 =2\,$cm ($\lambda_2 = 4$~cm), this results in an increase in the error of $1.8$ ($1.5$) relative to what the error would be if we used the single phase estimator at $\lambda_1$. 
We parameterize the increase in noise power from de-dispersion relative to the (higher $S/N$) shortest-wavelength band by the factor ${\cal A}$ and include this in our error calculations, taking a fiducial value of ${\cal A} = 1.5$ (cf. eqn.~\ref{eqn:Sh}).

\section{Refraction and diffraction}
\label{sec:refraction}

Since Appendix~\ref{sec:dispersion} showed that dispersion (which scales as $\phi_{\rm disp} \propto \lambda$) can be perfectly removed, the next question is how good of an assumption is it that $\phi_{\rm disp} \propto \lambda$?  For a homogeneous plasma, the correction to this long-wavelength scaling of the dispersion relation is suppressed by the ratio of the plasma frequency to the frequency squared.  Since the plasma frequency for interplanetary gas is very small ($\sim 10$ kHz) relative to the radio-wave frequencies of interest ($\sim 10$ GHz), this correction to the $\phi_{\rm disp} \propto \lambda$  scaling is extremely small.  

A potentially larger effect is that the radio waves at different frequencies travel different paths owing to diffraction or refraction caused by the density inhomogeneities in the solar wind.  Diffraction -- or multipath propagation -- results from phase fluctuations below the Fresnel scale $\sim \sqrt{\lambda L/(4\pi)}$.  The Fresnel scale physically corresponds to the impact parameter that encapsulates most light paths that contribute constructively to the image.  For the case at hand, the Fresnel scale corresponds to perturbations in the solar wind that pass the spacecraft with frequencies of $\omega_F  \sim v_{\rm sw}/\sqrt{\lambda L/(4\pi)} \sim 10\, (L/30 {\rm \,AU} \times \lambda/1 \, {\rm cm})^{1/2}~$Hz.  We can use our estimate for the RMS phase fluctuations (eqn.~\ref{eqn:phidisp}) to evaluate the phase fluctuations at $\omega_F$.  Evaluating at the fiducial parameters where the parentheses in equation~(\ref{eqn:hdisphrms}) are unity, the phase fluctuations at $ f_F \equiv \omega_F/(2\pi)$ are $f_F^{1/2} \widetilde{\phi}_{\rm disp, rms} \sim 10^{-7}$, and the leading-order contribution of diffraction to the RMS phase scales quadratically in $f_F^{1/2} \widetilde{\phi}_{\rm disp, rms}$ (as the linear order cancels since both positive and negative sub-Fresnel phase fluctuations contribute to the total).  This suggests that the RMS diffractive phase fluctuations should have an absolutely negligible value of $\sim (f_F^{1/2} \widetilde{\phi}_{\rm disp, rms})^2 \sim 10^{-14}$.  Furthermore, the sub-Fresnel fluctuations that drive the diffractive phase perturbations will decorrelate on timescales of $\omega_F^{-1}$, leading to further suppression over this estimate.  Thus, phase fluctuations from diffraction are likely to be extremely small.

Refraction is the displacement of images from a super-Fresnel density gradient. Refraction will contribute a larger effect since the phase perturbations in the solar wind are larger for $\omega \ll \omega_F$, and the longer wavenumbers of the participating modes mean there will be less decorrelation compared to diffraction.  However, in the following, we show that refraction also sources a negligible amount of phase noise.

To analyze the effect of refraction, let us assume that all phase is acquired halfway between the two spacecraft in an arm, an assumption that will put an upper bound on refractive-phase fluctuations.  The electric field's phase received at one receiver at wavelength $\lambda$ is given by the Fresnel-Kirchhoff integral (e.g.~\citealt{1992RSPTA.341..151N}) 
\begin{eqnarray}
\phi_\lambda &=& \arg \left( \frac{1}{2\pi i r_F^2} \int d^2\bfx \exp\left[i\frac{\bfx^2}{2 r_F^2} + i \phi_{d}(\lambda_j, \bfx) \right] \right).
\label{eqn:phij}
\end{eqnarray}
The integral is over the plane halfway between the two spacecraft, $\phi_{d}$ is the phase acquired by a sightline at position $\bfx$, and $r_F^2 \equiv \lambda L/(4\pi)$.  

Since refraction owes to the smooth contribution from longer wavelengths than $r_F$, we can approximate the phase with a quadratic function given by $\phi_{d}(\bfx)  \approx \phi_{d, l} + \nabla_i \phi_{d,l} x_i +  \frac{1}{2}\nabla_{i}\nabla_j \phi_{d,l} x_i x_j$, where all the Taylor expansion coefficients are evaluated at $\bfx = 0$.  Equation~(\ref{eqn:phij}) then evaluates to
\begin{eqnarray}
\phi_\lambda &\subseteq& -\frac{r_F(\lambda)^2}{2} \left (\frac{\nabla_i \phi_{d,l}(\bfx)}{1 + r_F(\lambda)^2 \nabla_i^2 \phi_{d,l}(\bfx)} \right)^2, \nonumber \\ 
&=& -\frac{r_F(\lambda)^2}{2} \Big|\boldsymbol{\nabla} \phi_{d,l}(\bfx)\Big|^2 + \frac{r_F(\lambda)^4}{2} \Big(\nabla_i \phi_{d,l}(\bfx)\Big)^2  \nabla_i^2 \phi_{d,l}(\bfx) +  ... \ , 
\label{eqn:expansion}
\end{eqnarray}
where we have reconstituted the $\bfx$ argument for the coefficients in the expansion so that we can capture their time variability as the solar wind flows past, as well as omitted the wavelength argument in $\phi_{d}(\lambda_j, \bfx)$ to simplify notation, and we have dropped an overall constant, as this term is what was considered in Appendix~\ref{sec:dispersion} and is perfectly removed by the de-dispersion estimator given by equation~(\ref{eqn:dedisp}).  We have rotated to the basis where $\nabla_{i}\nabla_j \phi_{d,l}(0)$ is diagonal, and $i$ indices are implicitly summed.  Equation~(\ref{eqn:expansion}) shows that the leading refractive term is the square of the phase gradient times the Fresnel scale.  Since each $r_F \nabla$ brings in a factor of $f/f_F$ where $f_F \equiv \omega_F/(2\pi)$, the refractive effect should be suppressed at $f \sim 10^{-5}\;$Hz and $\lambda = 1~$cm (noting that then $f/f_F\sim 10^{-6}$) by a factor of $\sim 10^{-13}$ relative to the mean dispersive term, which is $\propto \phi_{d, l}$ and whose effect was estimated in the previous section.  This order-of-magnitude estimate uses that the phase fluctuation at $f= 10^{-5}\;$Hz is $\sim 0.1$.  This simple estimate is consistent with the following more detailed calculation.

To calculate the phase noise from refraction in more detail, let us consider the lowest-order non-constant term in equation~(\ref{eqn:expansion}).  Assuming the same setup as in Appendix~\ref{sec:dispersion} and illustrated in Figure~\ref{fig:dispersionvis}, where temporal correlations can be related to spatial correlations of the advecting solar wind, we can calculate the corrections from refraction owing to phase correlations as
\begin{eqnarray}
\Big\langle\phi_\lambda(t)  \phi_\lambda(t +\tau)\Big\rangle \ &\approx& \  r_F(\lambda)^4   \Big\langle\big|\boldsymbol{\nabla} \phi_{d,l}(\bfx)\big|^2 \ \big| \boldsymbol{\nabla} \phi_{d,l}(\bfx + v_{\rm sw} \tau \hatn)\big|^2\Big\rangle  \nonumber \\
    &\subseteq& \  2 r_F(\lambda)^4 \xi_{\nabla \phi}(\tau)^2, \label{eqn:refractphasecorr}
\end{eqnarray}
where in addition to dropping an irrelevant constant, the last line assumed statistical isotropy and Gaussian statistics to express the four-point correlation function in terms of the two-point correlation function, where
$$ \xi_{\nabla \phi}(\tau) = \Big\langle{\boldsymbol{\nabla} \phi_{d,l}(\bfx) \cdot \boldsymbol{\nabla} \phi_{d,l}(\bfx + v_{\rm sw} \tau \hatn)\Big\rangle}.$$
The half-bandwidth power spectrum of phase fluctuations from refraction can be calculated from the Fourier transform of equation~(\ref{eqn:refractphasecorr}):
\begin{equation}
P_{\rm \phi, ref}(f = 2\pi \omega) =2 r_F(\lambda)^4 \int dt e^{-i \omega t} \xi_{\nabla \phi}(t)^2 = 2 r_F(\lambda)^4 \int \frac{d\omega'}{2\pi} P_{\nabla \phi}(\omega') P_{\nabla \phi}(\omega - \omega'), 
\label{eqn:Pphiref}
\end{equation}
where the full-bandwidth power spectrum of $|\boldsymbol{\nabla}  \phi|$ is defined as
\begin{eqnarray}
P_{\nabla \phi}(\omega) &\equiv & \int dt e^{-i \omega t}  \xi_{\nabla \phi}(t) 
=    \frac{ (2\pi c)^2}{\lambda^2}\int dt e^{-i \omega t} \expectation{\nabla_i \Delta\tau_{d,1} \nabla_i \Delta\tau_{d,2}}(v_{\rm sw} t),  \label{eqn:Pgradphi}
\end{eqnarray}
and we have converted to gradients of the time delay rather than phase using that $\phi_d = 2\pi c/\lambda \,\tau_d$.  Using the integral expression given by equation~(\ref{eqn:vartauintegral}) for $\expectation{\Delta\tau_{d,1}  \Delta\tau_{d,2}}$, and that the derivatives bring down factors of $\bfk_\perp$ under the integrand, we can write equation~(\ref{eqn:Pgradphi}) as
\begin{eqnarray}
P_{\nabla \phi}(\omega) &=&  \int dt e^{-i \omega t}   \kappa'^2 L \int \frac{d^2\bfk_\perp}{(2\pi)^2} P_e(k) \Big|\bfk_\perp\Big|^2 e^{i \bfk_\perp \cdot  \hatn  v_{\rm sw} t},\\
&=& \frac{\kappa'^2 L}{v_{\rm sw}} \int_{-\infty}^{\infty} \frac{d k_x}{(2\pi)} P_e\left(\sqrt{k_x^2+  \frac{\omega^2}{v_{\rm sw}^2}} \right) \left(k_x^2+ \frac{\omega^2}{v_{\rm sw}^2}  \right),\\
&\approx & \frac{4.4 \kappa'^2 L |\omega|}{ v_{\rm sw}} P_e^{\rm 1D}(f, r),
\end{eqnarray}
where $\kappa' = 2 \pi c \kappa/\lambda$ and, similarly to how we reached equation~(\ref{eqn:Ptaufinal}), we assumed a power-law for the electron density power spectrum. (The coefficient changes to 3.1 if $\beta = 2$ is taken for the spectral slope of the 1D power spectrum rather than $\beta = 1.5$, as was assumed to evaluate this expression.) The convolution that yields the phase noise power  $P_{\rm \phi, ref}$ (eqn.~\ref{eqn:Pphiref}) is only convergent for $3/2 < \beta <2$, which conveniently is the rough range of indices favored by the Voyager 2 data, and has the scaling
\begin{equation}
 P_{\rm \phi, ref}(f) \sim  10^2 \frac{r_F(\lambda)^4 \kappa'^4 L^2 f}{v_{\rm sw}^2} \Big[f P_e^{1D}(f, r) \Big]^2 =   10^2 \frac{\pi^2 \kappa^4 L^4 f}{\lambda^2 v_{\rm sw}^2} \Big[f P_e^{\rm 1D}(f, r) \Big]^2,
\end{equation}
where we have stuck in a $10^2$ prefactor to match our order-of-magnitude estimates from evaluating this integrand.

Finally, our expression for the RMS strain error using this estimate for $ P_{\rm \phi, ref}$ and our de-dispersion estimator given in equation~(\ref{eqn:dedisp}) becomes 
\begin{eqnarray}
\widetilde{h}_{\rm ref, rms} \ &=& \ f(\lambda_1, \lambda_2) \frac{\lambda_1 \sqrt{4 P_{\rm \phi, ref}(f)|_{\lambda_1}}}{2\pi L  ~\sqrt{R(f)}}, \nonumber\\
&\ \sim & \ 10^{-25}~~ {\rm Hz}^{-1/2} ~R(f)^{-1/2} \left( \frac{f(\lambda_1, \lambda_2)}{4}\right) \left( \frac{\lambda_1}{1 \,\rm cm} \right)^4  \left(\frac{f}{10^{-5} {\rm Hz}}  \right)^{(3-2\beta)/2} , \nonumber \\
& & \ \times\left(\frac{ P_{e,0}^{\rm 1D}}{10^{11.1}~\text{m$^{-6}$ Hz$^{-1}$}} \right) \left( \frac{r}{30 \, \rm AU} \right)^{-3} \left( \frac{L}{30 \, \rm AU}\right)^2.\label{eqn:href_eval}
\end{eqnarray}
where $f(\lambda_1, \lambda_2) = \Big|\frac{ \lambda_2/\lambda_1^3}{(\lambda_1^2 - 
\lambda_2^2)}\left(\lambda_1 \lambda_2^3 -\lambda_2\lambda_1^3\right)\Big|$, which evaluates to $f(\lambda_1, \lambda_2) = 4$ if $\lambda_2 = 2\lambda_1$ and we have used the parameterization of the power given by  equation~(\ref{eqn:swdensitypower}).  Equation~(\ref{eqn:href_eval}) shows that the effect of refraction is very small relative to other sources of noise.  

\section{Data rates}
\label{sec:datarates}

Due to the tens of astronomical units distances of the spacecraft in our concepts, the concepts developed here ire likely constrained to a downlink data rate of $\sim 10$~kbps as has been achieved by previous outer Solar System missions such as the New Horizons spacecraft \citep{2008SSRv..140...23F}.
To estimate how much data would need to be telemetered back to Earth, let us compare the proposed mission with the LISA mission.  LISA aims to send a downlink with a data rate of $230\;$kbps for eight hours every day \citep{LISA}. 
Since LISA's goal is to constrain gravitational waves to frequencies as high as $1$~Hz, it is designed to sample at 4~Hz \citep{PhysRevD.107.083019, LISA}.  Because LISA has three interferometric arms, the downlinked data include data streams from the six interferometric observables as well as from the positions of six test masses.

The proposed mission likely could accommodate a total data transmission that is a factor of $\sim10^5$ smaller than LISA because (1) it targets $f < 10^{-4}$Hz and so can sample on $10^4\times$ longer timescales; (2) it has one or two arms and no test masses, such that there are fewer data streams to downlink; and (3) it would have $\sim 10^4$ times less precision in the phase measurements relative to LISA and, hence, require fewer bits per sample.  This much lower data rate means that a one-hour downlink at 10~kbps could potentially download an entire year of science data. 

\section{Angular resolution of an arm}
\label{ap:angres}


This appendix justifies the claim in \S~\ref{ss:localization} that the angular resolution of a single arm of our gravitational wave concepts is $\delta \theta \sim \lambda_{\rm GW}/(L~ {\rm SNR})$ where $\theta$ is the polar angle defined by the arm, as long as the source drifts sufficiently in frequency.  
We further derive a condition for the total amount of frequency drift for this expression to hold. 

To start, the Doppler shift imparted by a passing gravitational wave with aligned polarization at one of the phase readouts is \citep{1975GReGr...6..439E}
\begin{equation}
\frac{\delta\nu}{\nu} = \frac{1}{2}\left[(1-\sin\theta)h(t) + 2\sin\theta \, h(t-L/c-L/c\sin\theta) - (1+\sin\theta)h(t-2L/c)\right],
\end{equation}
where $\theta = 0$ corresponds to perpendicular propagation relative to the arm. The middle term contains the phase-dependent information across the arm that can be used for localization, with this phase information arising from the time delay of the gravitational wave signal as it propagates across the arm. For small angular deviations $\delta\theta$ from some reference angle $\theta_*$, this term modulates the gravitational wave with phase:
\begin{equation}
\phi = \frac{2\pi L \delta\theta}{\lambda_{\rm GW}} \cos(\theta_*).
\end{equation}

Matched-filtering detection of gravitational wave strain allows phase determination to precision $\delta\phi \sim 1/{\rm SNR}$, where SNR is the signal-to-noise ratio of the gravitational wave detection.  Equating the phase uncertainty to the geometric phase difference:
\begin{equation}
\frac{1}{\rm SNR} \sim \delta\phi = \frac{2\pi \, L \delta\theta}{\lambda_{\rm GW}} \, \cos(\theta_*),
\end{equation}
implying an angular uncertainty in the polar angle with respect to the arm of
\begin{equation}
\delta\theta \sim \frac{\lambda_{\rm GW} \cos(\theta_*)}{2\pi \,L\, {\rm SNR}},
\label{eqn:deltathetaderiv}
\end{equation}
which, up to the factor of $\cos(\theta_*)/2\pi$, equals our ballpark estimate.

The derivation to this point does not consider that for wavelengths smaller than the arm length ($\lambda_{\rm GW} < L$), a single arm cannot distinguish a monochromatic wave that in projection spans an integer number of wavelengths across the arm such that
\begin{equation}
\frac{L\sin\theta}{\lambda_{\rm GW}} = n,
\end{equation}
for integer $n$.  Thus, for a monochromatic wave, equation~(\ref{eqn:deltathetaderiv}) represents the localization precision around \emph{each} $n$. This angular degeneracy can be broken by observing the source across multiple frequencies. Consider two frequency bins, $f_1$ and $f_2$, separated by a bandwidth of $B = f_2 - f_1$. To distinguish between the $n=0$ and $n=1$ solutions, the angular error $\delta \theta$ must be smaller than the difference between solutions at the two frequencies. In the small-angle approximation, this difference is
\begin{equation}
 \delta \theta_{\Delta n = 1} = \frac{c}{f_1 L} - \frac{c}{f_2 L} \approx \frac{c B}{f^2 L},
\end{equation}
where $f$ is the average frequency and the last expression is valid to the extent $B/f \ll 1$. This separation must be larger than our angular resolution in order to choose the correct $n$, which is the condition that
\begin{equation}
 \delta \theta_{\Delta n = 1} > \delta \theta ~~~\longrightarrow~~~
\frac{B}{f} > \frac{1}{\rm SNR}.
\end{equation}


We find our approximate expression $\delta \theta \sim \lambda_{\rm GW}/(L~ {\rm SNR})$ provides decent estimates for the localization performance of LISA relative to the more detailed estimates in \citet{2021PhRvD.103h3011M} for a binary merger at $2.5~$ hours and $7~$ min from coalescence (see their Figure 13).  It also applies to the concepts presented in this paper.


\end{document}